%% file: paper_v3.tex
\documentclass[a4paper,conference]{IEEEtran}
\usepackage{graphicx}
\usepackage{algorithm,algorithmic}
\usepackage{setspace}
\usepackage[font=scriptsize]{subcaption}
\usepackage[font=footnotesize]{caption}
\usepackage{mathtools}
\usepackage{amsthm}
\usepackage[draft]{hyperref} 
\usepackage{cite}
\usepackage{amssymb}
\usepackage{physics}
\usepackage{booktabs} 

\usepackage{tikz}
\usepackage{pgfplots}
\usepackage{tikzscale}
\usepackage{tikz-qtree}
\input{myTikz.tex}

\usepackage[nowarn,acronyms,nonumberlist,nopostdot,nomain,nogroupskip]{glossaries}
\usepackage{xcolor}

\makeatletter
\newcommand{\LeftEqNo}{\let\veqno\@@leqno}
\makeatother

\input{acronyms.tex}
\makeglossaries
\glsdisablehyper

\newcommand{\AZ}[1]{}

\usepackage[capitalize]{cleveref}
\crefname{section}{Sec.}{Secs.}

\IEEEoverridecommandlockouts
\begin{document}


\title{A Simulation Framework for\\Contention-Free Scheduling on WiGig}

\author{
 \IEEEauthorblockN{Matteo Drago$^*$, Tommy Azzino$^{\circ}$, Mattia Lecci$^*$, Andrea Zanella$^*$, Michele Zorzi$^*$}
 \IEEEauthorblockA{$^*$\textit{Department of Information Engineering}, \textit{University of Padova}, Italy, E-mails: \texttt{\{name.surname\}@dei.unipd.it}}
  \IEEEauthorblockA{$^{\circ}$\textit{NYU Tandon School of Engineering}, Brooklyn, NY, USA, E-mail: \texttt{tommy.azzino@nyu.edu}}
 \thanks{This work was partially supported by NIST under Award No. 60NANB19D122.
  Mattia Lecci's activities were supported by \textit{Fondazione CaRiPaRo} under the grant ``Dottorati di Ricerca 2018.''}
}

\maketitle


\glsunset{wifi}

\begin{abstract}
The latest IEEE 802.11 amendments provide support to directional communications in the \acrlong{mmw} spectrum and, thanks to the wide bandwidth available at such frequencies, makes it possible to wirelessly approach several emergent use cases, such as virtual and augmented reality, telepresence, and remote control of industrial facilities.
However, these applications require stringent \acrlong{qos}, that only contention-free scheduling algorithms can guarantee.
In this paper, we propose an end-to-end framework for the joint admission control and scheduling of periodic traffic streams over mmWave \acrlongpl{wlan} based on \acrlong{ns3}, a popular full-stack open-source network simulator.
Moreover, we design a baseline algorithm to handle scheduling requests, and we evaluate its performance with a full-stack perspective.
The algorithm is tested in three scenarios, where we investigated different configurations and features to highlight the differences and trade-offs between contention-based and contention-free access strategies.


\end{abstract}


\IEEEpeerreviewmaketitle

\begin{tikzpicture}[remember picture,overlay]
\node[anchor=north,yshift=-20pt] at (current page.north) {\parbox{\dimexpr\textwidth-\fboxsep-\fboxrule\relax}{
\centering\footnotesize This paper has been submitted to IEEE GLOBECOM 2021. Copyright may change without notice.}};
\end{tikzpicture}

\glsresetall
\glsunset{wifi}

\section{Introduction} 
\label{sec:introduction}

Indoor \gls{wifi} networks have had a key role in the digital revolution of the last two decades, as wireless technologies paved the way toward the design of applications for work settings (e.g., smart metering, remote control) and house entertainment (e.g., \gls{ar}, \gls{vr}, \gls{xr}).
From a technical point of view, these new applications also changed the infrastructure requirements, with higher required data rate, lower delay thresholds, and brand new classes of \gls{qos} constraints.

To face these challenges, moving to the \gls{mmw} spectrum has proven to be a valuable alternative to the widespread sub-6~GHz spectrum used by legacy wireless architectures, given the abundant bandwidth available in the former frequency range.
For this reason, in an effort to create a common playground for researchers and manufacturers, the IEEE devised specific amendments to update the \gls{phy} and \gls{mac} layers in what is known as \gls{wigig}, first with 802.11ad~\cite{standard802.11ad} in 2012 and now with 802.11ay~\cite{tgayWebsite}.

In particular, \gls{wigig} standards introduced a new contention-free strategy to access the transmission medium at specific time intervals, referred to as \glspl{sp}. A \gls{sta} can request \glspl{sp} to the \gls{pcpap} asking for a specific duration and periodicity.
A detailed overview of such procedure will be later described in \cref{sec:ns_3_scheduling_framework}.

This new access strategy can be useful for applications with stringent \gls{qos} requirements, i.e., throughput, delay, and jitter, which may be heavily affected by legacy, contention-based channel access mechanisms.
Moreover, applications such as video streaming or \gls{vr} can generate periodic traffic, whose performance with contention-based channel access can degrade, given the uncertain availability of resources from one time interval to another.
Fortunately, \gls{wigig} provides specific scheduling mechanisms to directly support periodic applications with tight \gls{qos} constraints.
From a practical point of view, however, dealing with concurrent periodic traffic streams from multiple users is not easy since the designed policy should be able to manage heterogeneous requests, while possibly guaranteeing fairness among different flows.

Considering all these aspects, in this work we propose an \gls{e2e} framework to manage distinct traffic flows based on the requirements provided by the \gls{wigig} standards, taking care of the admission and scheduling of new allocation requests.
Besides, using this framework, we design a baseline algorithm to allocate periodic requests and validate its trade-offs through a detailed full-stack performance evaluation.
To do so, we extend the module described in \cite{assasa2016implementation}, which integrates into \gls{ns3} the new features of 802.11ad, and publicly release the source code to the research community.

The rest of the paper is organized as follows.
In \cref{sec:state_of_the_art} we overview the literature on this topic, while in \cref{sec:ns_3_scheduling_framework} we describe the framework we designed and implemented in an open-source full-stack network simulator.
Then, we accurately describe the simulation setup in \cref{sec:simulation_setup} and discuss the performance of the scheduling algorithms in \cref{sec:results}.
Lastly, in \cref{sec:conclusions} we draw our conclusions and propose possible extensions to this work.


\section{State of the Art} 
\label{sec:state_of_the_art}

The optimization of \gls{wifi}'s \gls{mac} layer procedures has been investigated in the literature, even before \gls{wigig} standards were introduced.
Most of these works, however, mainly focus on \glspl{cbap} and do not consider the possibility of using Service Periods (SPs).
Starting from 802.11ad, the possibility of allocating contention-free resources gained further momentum, considering also the directional characteristic of mmWave channels.
An attempt to prioritize the traffic injected in the network was done for 802.11e, where four \glspl{ac} were introduced. 
Based on which category they belong to, packets with higher priority use a shorter \gls{aifs} and thus they wait less before being transmitted.
A study of 802.11e contention-based prioritization mechanisms was provided in \cite{bianchi2005contentionbased}.

A mathematical framework to analyze \gls{e2e} metrics on 802.11-based systems was proposed in \cite{babich2009throughput}, to compare throughput and average packet delay in scenarios where the nodes are equipped with advanced antenna systems. 
It also accounts in detail for the characteristics of the \gls{dcf}, for which a theoretical performance analysis was carried out in \cite{bianchi2000performance}.

Likewise, the authors in~\cite{pielli2019analytical} presented a detailed analytical model to assess the performance of \glspl{cbap} in 802.11ad, taking into account a directional channel model and the presence of scheduled \glspl{sp}.
Yet, the model lacked the details about how to schedule such \glspl{sp} for certain types of relevant applications, such as periodic ones.

A seminal study on the use of \gls{rl} to solve the problem of jointly scheduling \glspl{cbap} and \glspl{sp} in 802.11ad is in~\cite{azzino2020scheduling}, where \gls{ns3} was used to assess how the algorithm could decrease the \gls{dti} occupancy while guaranteeing state-of-the-art \gls{qos} performance.

In general, in the context of \gls{wigig} networks, little work has been done on the scheduling of contention-free time resources.
Moreover, to the best of our knowledge, little to no work in the literature faces the problem of periodic scheduling with all the constraints introduced by \gls{wigig} standards.
The authors of~\cite{hemanth2013performance,rajan2016saturation}, for example, study the case where all \glspl{sp} are allocated at the beginning of each \gls{bi}, while the rest of the interval is left for a single \gls{cbap}.
In~\cite{khorov2016mathematical} they propose an accurate mathematical analysis of the performance of a realistic \gls{vbr} traffic source in the presence of channel errors when using a periodic resource allocation scheme.
How to schedule multiple allocations at once, however, has not been detailed.

On the other hand, the problem of periodic scheduling has been studied for real-time computation and task scheduling, where the goal is to complete some tasks before a certain deadline while minimizing resource utilization.
For example, the authors of~\cite{liu73scheduling} proposed a scheduling algorithm to dynamically assign priorities, capable of achieving full processor utilization.
In~\cite{ramamritham95allocation}, the authors provide a framework for allocating periodic tasks in multiprocessor systems, which takes into account their requirements, while periodicity constraints are translated into time deadlines.

All these approaches, however, cannot be adapted to the \gls{wigig} framework, as the constraints imposed by the standardized resource allocation procedures are completely different.


\section{ns-3 Scheduling Framework} 
\label{sec:ns_3_scheduling_framework}


In this section, we describe the design choices and assumptions necessary to implement our scheduling framework on top of the 802.11ad \gls{ns3} module~\cite{assasa2016implementation}, with a focus on \gls{mac} layer mechanisms.

First of all, \gls{wigig} standards refer to the \gls{bi} as the unit time interval used by the devices to organize association, beamforming, and data transmission procedures.
To this aim, it is further divided into \gls{bhi} and \gls{dti}.

The \gls{bhi} is then organized into \gls{bti}, \gls{abft}, and \gls{ati}, and it is devoted to association, beamforming, and scheduling procedures.
\glspl{sta} can communicate during the \gls{dti} using \acrfullpl{cbap}, or using dedicated and prearranged \glspl{sp}, that guarantee some resources to a given user who made request to the \gls{pcpap}.


The standards do not pose any constraint on the number, order, or type of these transmission slots in the \gls{dti}, however, each \gls{sta} must follow a common procedure to request such resources.
Since we are mainly interested in periodic applications, in this work we focus on \textit{isochronous pseudo-static} allocations allocated using \gls{addts} Request/Response scheduling elements. 

First, a \gls{sta} sends an \gls{addts} Request to the \gls{pcpap}, specifying parameters such as the \textit{Allocation Period} (if any), the \textit{Minimum} and \textit{Maximum Allocation} duration in each allocation period, the \textit{pseudo-static} flag, which allows for persistent allocations over multiple consecutive \glspl{bi}, among others.
After that, if the \gls{pcpap} can accommodate the new request, it sends back an \gls{addts} Response, specifying the allocated duration and the starting time.


Our work~\cite{ns3-802.11ad-scheduling} mainly focused on the design and implementation of a generic scheduling interfaces, called \texttt{DmgWifiScheduler}, that implements the scheduling features for the \gls{mac} entity of the \gls{pcpap}.
Starting from this class, we extended it to create the \texttt{PeriodicWifiScheduler}, a simple scheduler for the allocation of periodic resources.
Moreover, to study how a contention-based-only approach affects the overall \gls{qos}, we also created the \texttt{CbapOnlyWifiScheduler}, forcing STAs to transmit only over CBAP by allocating the entire \gls{dti} as such.

Even though the performance evaluation, presented in \cref{sec:results}, considers only allocations with the same period and application requirements for all \glspl{sta}, scheduled starting from the beginning of the DTI back to back as long as they fit, it is crucial to elaborate on the design choices that lead to this framework.

Thus, \texttt{PeriodicWifiScheduler} includes the following assumptions:
\begin{itemize}
  \item Only SP allocations with period equal to an integer fraction of a \gls{bi} are supported, while the standards also support periods multiple of the \gls{bi}.
  \item If the period is $t=T_{BI}/p$, the request is accepted only if the available time in the \gls{dti} can accommodate exactly \textit{p} \glspl{sp}, commonly referred to as allocation blocks, each distanced by $t$.
  For example, if \textit{p} $=4$, the number of blocks per \gls{bi} must be exactly $4$.
  \item A \gls{sta} can send an \gls{addts} Request to reduce the duration of the allocation, while the increase is not supported as it possibly requires a major reorganization of the \gls{dti}.
  \item Once an allocation is accepted, the \glspl{sp} duration and blocks starting time cannot be changed by the scheduler, even if the \gls{dti} structure changes as a consequence of subsequent requests from other \glspl{sta}.
  \item All the time that is not reserved by \glspl{sp} will be allocated as \gls{cbap}.
\end{itemize}

These constraints allowed us to validate our results in a clear setting with firm requirements.

\section{Simulation Setup} 
\label{sec:simulation_setup}

\begin{table}[t!]
\caption{Simulation parameters}
\label{tab:params}
\footnotesize
\centering
\begin{tabular}{ll|ll}
\toprule
MCS                    & $4$ (fixed)  & APP period ($T_{APP}$) & $T$        \\
Max A-MSDU size        & $7\,935$ B   & Packet size            & $1\,448$ B \\
Max A-MPDU size        & $262\,143$ B & Traffic direction      & Uplink     \\
BI duration ($T_{BI}$) & $T$          & Simulation duration    & $10$ s     \\
SP period ($T_{SP}$)   & $T$          & Independent runs       & 30         \\
Network protocols      & IPv4/UDP     & $T$                    & $102.4$ ms \\
\bottomrule
\end{tabular}
\end{table}


The network scenario consists of a single \gls{pcpap} in the center of a room, surrounded by \glspl{sta} with perfect channel conditions, with simulation parameters listed in \cref{tab:params}. 
%

To emulate periodic traffic, we implemented a \textit{periodic application} that generates periodic packet bursts, whose size and period can be set as a parameter of the application, with every single packet being of size $1\,448$~B.
Traffic is generated by the \glspl{sta} and sent to the \gls{pcpap}.

Since we expect CBAP-only scheduling to yield good performance when a small amount of traffic is sent over the network, and the SP scheduling to show its full potential for highly loaded networks, we defined the \textit{normalized offered traffic} which we refer to as $\eta$.
By varying $\eta$ in $(0, 1]$, we control the traffic injected in the network, equally distributed among the number of stations.

For instance, in a scenario with $N=4$~\glspl{sta} transmitting using \gls{mcs}~$4$ with a nominal PHY rate of $R_4=1\,155$~Mbps, for $\eta=0.5$ the aggregate average offered traffic should be $\eta R_4=577.5$~Mbps, and thus each STA will generate about $\eta R_4/N\approx 144$~Mbps.

Note that with $\eta=1$, the offered PHY rate would be exactly $1\,155$~Mbps, thus overloading the network.
In fact, a portion of each BI is always reserved for the \gls{bhi} where STAs are not allowed to transmit information, reducing the overall network capacity.
On the other hand, $\eta=0$ would translate in no traffic injected into the network.
For this reason, in \cref{sec:results} we will show results for traffic loads $\eta\in[0.01, 0.9]$.

In all our simulations, the period of all periodic applications $T_{APP}$, the period of all scheduled SPs $T_{SP}$, and the duration of the BI $T_{BI}$ are all the same, and thus simply noted as $T=102.4$~ms.
Based on the value of $\eta$, the number of packets making up a burst is constant as well, and they are all generated at the beginning of each application period.

The duration of each SP is computed based on the MCS and the application rate for the full transmission burst to fit exactly into the SP.
The minimum and maximum duration fields in the \gls{addts} Request are thus equal, meaning that the request is either accepted by the \gls{pcpap} guaranteeing the exact amount of resources necessary to serve its application, or rejected, and the rejected STA will remain silent for all the simulation.

If the \gls{addts} Response for a given STA is accepted, its application will start randomly over a period $T$, and thus, by default, will not be aligned with the beginning of its assigned \glspl{sp}. 
To fully take advantage of the scheduling concept, however, application and \glspl{sp} should be aligned to yield the best possible performance.
To do so, the APP layer has to be aware that the transmission will happen over a \gls{wigig} network as well as the details of the scheduled \glspl{sp}, requiring some information exchange with the MAC layer.
This might be possible for some types of applications running on specific hardware, e.g., \gls{vr} headsets and, in general, for high-end hardware running applications that require tight delay constraints.
For this reason, we defined a \textit{smart mode} which, if activated, makes the application start at the beginning of the first allocated \gls{sp}, thus assuming a cross-layer interaction and alignment.
Nonetheless, this does not take into account applications with non-deterministic periods, which could lose the alignment in the following \glspl{sp}.

\begin{figure*}[t!]
  \begin{subfigure}[b]{\textwidth}
   \centering
   \input{img/legend.tex}
  \end{subfigure}
 \\
 \newcommand\fheight{0.7\columnwidth}
 \newcommand\fwidth{1\columnwidth}
  \hspace*{\fill}%
  \begin{subfigure}[b]{0.3\textwidth}
   \centering
   \input{img/burst/avg_delay.tex}
   \caption{Average delay}
   \label{fig:burst_avg_delay}
  \end{subfigure}
  \hspace*{\fill}%
  \begin{subfigure}[b]{0.3\textwidth}
   \centering
   \input{img/burst/avg_delay_variation.tex}
   \caption{Average delay variation (jitter)}
   \label{fig:burst_avg_delay_variation}
  \end{subfigure}
  \hspace*{\fill}%
  \begin{subfigure}[b]{0.3\textwidth}
   \centering
   \input{img/burst/norm_thr.tex}
   \caption{Normalized aggregated throughput}
   \label{fig:burst_norm_thr}
  \end{subfigure}
  \hspace*{\fill}%
 
  \caption{Performance of the different scheduling configurations with a bursty application with deterministic period $T=102.4$~ms.}
  \label{fig:burst}
 \end{figure*}
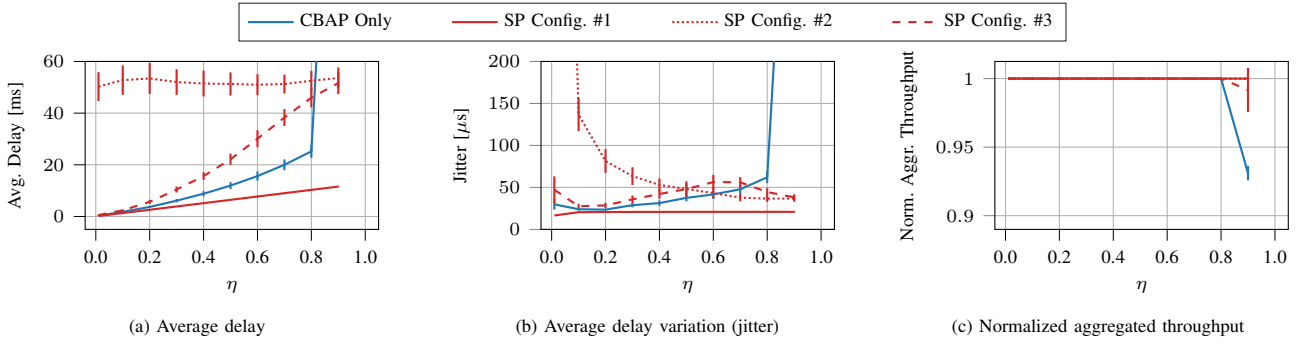

We compare the performance of four scheduling configurations, namely:
\begin{itemize}
  \item \textit{CBAP-only}: all STAs transmit during the CBAP.
  \item \textit{SP Config. \#1}: the \textit{smart start} mode is enabled.
  STAs are also allowed to transmit in the \gls{cbap} if necessary.
  \item \textit{SP Config. \#2}: \textit{smart start} is disabled and \glspl{sta} cannot transmit in the \gls{cbap}.
  \item \textit{SP Config. \#3}: \textit{smart start} is disabled and \glspl{sta} are allowed to transmit in the \gls{cbap} if necessary.
\end{itemize}

The performance evaluation of the proposed scheduling schemes has been carried out in three distinct scenarios.
\begin{itemize}
  \item \textit{First scenario}: four STAs transmit at different values of $\eta$ using a deterministic application with period $T$.
  \item \textit{Second scenario}: all applications offer the same APP-layer rate of $R=50$, $100$, $200$~Mbps with a deterministic period of $T$, varying number of STAs up to 10.
  \item \textit{Third scenario}: four STAs transmit a heavy traffic load ($\eta=0.75$) using applications with random period. Periods are independently sampled one after the other $\mathbf{T}_i=\mathcal{N}(\mu, \sigma^2)$, where $\mu = T$ and $\sigma=\rho T$, calling $\rho$ the \textit{period deviation ratio}. Thus, for a given STA, bursts will occur at times $\mathbf{t}_k = t_0 + \sum_{i=1}^k \mathbf{T}_i$.
\end{itemize}


\section{Results} 
\label{sec:results}

In this section, we evaluate the performance of the different configurations considering a number of packet-based \glspl{kpi}.
First of all, the \textit{average delay} takes into account only successfully received packets. 
For some relevant scenarios we also show the packet \textit{jitter}~\cite{rfc3393}, defined as the average absolute delay variation among successive packets.
The \textit{aggregated throughput} is also considered as a metric for network utilization, sometimes normalized by the amount of aggregated offered traffic.
Finally, all metrics also show the $95\%$ confidence intervals computed as $1.96 \frac{\sigma_{\rm runs}}{\sqrt{N_{\rm runs}}}$.

\paragraph{First Scenario} 
\label{par:first_scenario}

\cref{fig:burst} shows the results for the first proposed scenario, where we compare the four scheduling configurations against traffic load, considering a deterministic application, as described in \cref{sec:simulation_setup}.

In \cref{fig:burst_avg_delay} we show the \textit{average delay} for this scenario.
Note that an increasing $\eta$ directly translates into an increased burst size, since more packets have to be delivered in a given period~$T$, thus increasing the achievable average delay.

When the scheduling of \glspl{sp} is not allowed, \gls{cbap}-only offers almost ideal delay performance for low traffic loads, which however degrades for higher loads and even becomes unstable for $\eta>0.8$.

Instead, \gls{sp} configuration \#1, i.e., using \textit{smart start}, is clearly the optimal strategy and represents a lower bound for all other configurations, since packets are sent immediately and back-to-back.

SP configuration \#2, where \textit{smart start} is not used and STAs with scheduled SPs are \textit{not} allowed to access the CBAP, shows an almost constant average delay of about $51.2$~ms~$=T/2$.
It can be proven that an application with period $T$ with a uniformly distributed start time, which can only transmit during an SP of the same periodicity $T$ and with a duration equal to what is needed to transmit the packet burst, has an expected average delay of exactly $T/2$, irrespective of the traffic load or the number of transmitting nodes.
In fact, application bursts will happen either (i) sometime during the ongoing SP, so that the next SP will also be needed to finish sending the whole burst causing a large increase of the average delay, or (ii) outside an SP, thus needing to wait for the start of the next SP but being able to send the whole burst at once.

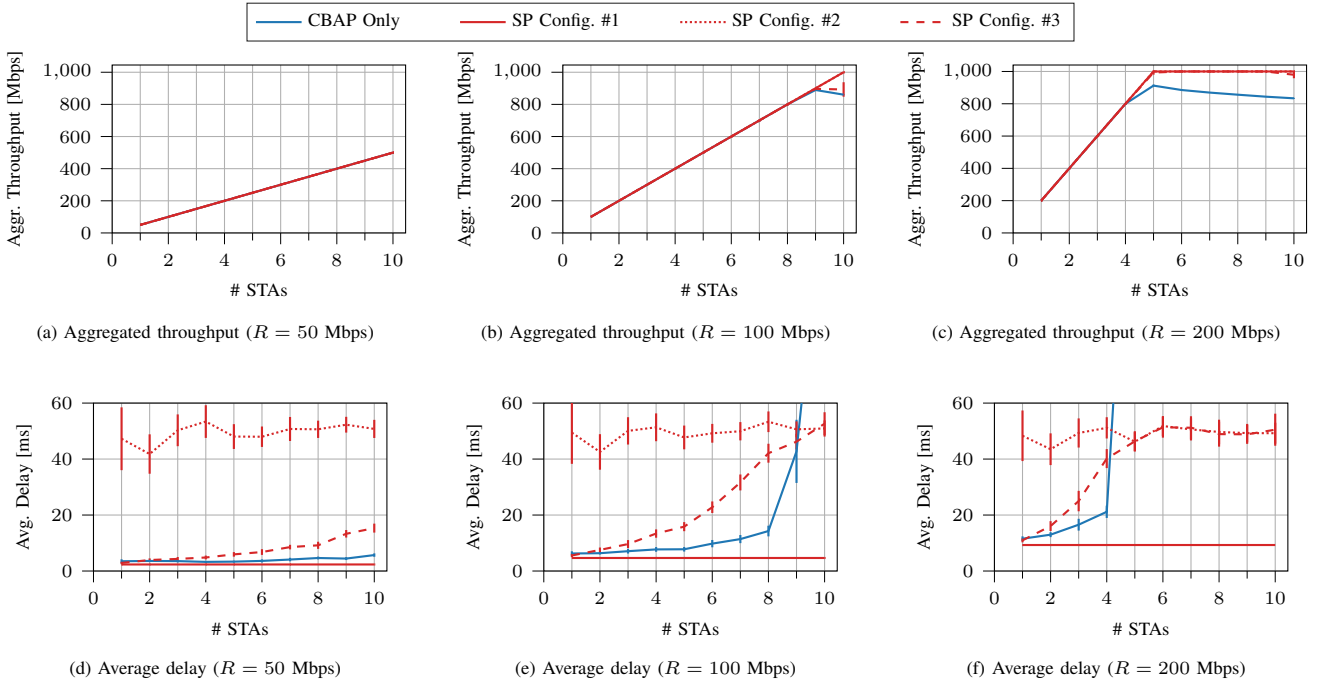
\begin{figure*}[t!]
  \begin{subfigure}[b]{\textwidth}
   \centering
   \input{img/legend.tex}
  \end{subfigure}
 \\
 \newcommand\fheight{0.7\columnwidth}
 \newcommand\fwidth{1\columnwidth}
  \hspace*{\fill}%
  \begin{subfigure}[b]{0.3\textwidth}
   \centering
   \input{img/50Mbps/aggr_thr.tex}
   \vspace{-2.5ex}
   \caption{Aggregated throughput ($R=50$~Mbps)}
   \label{fig:50Mbps_aggr_thr}
  \end{subfigure}
  \hspace*{\fill}%
  \begin{subfigure}[b]{0.3\textwidth}
   \centering
   \input{img/100Mbps/aggr_thr.tex}
   \vspace{-2.5ex}
   \caption{Aggregated throughput ($R=100$~Mbps)}
   \label{fig:100Mbps_aggr_thr}
  \end{subfigure}
  \hspace*{\fill}%
  \begin{subfigure}[b]{0.3\textwidth}
   \centering
   \input{img/200Mbps/aggr_thr.tex}
   \vspace{-2.5ex}
   \caption{Aggregated throughput ($R=200$~Mbps)}
   \label{fig:200Mbps_aggr_thr}
  \end{subfigure}
  \hspace*{\fill}%
  \\\vspace{1ex}
  \\
  \hspace*{\fill}%
  \begin{subfigure}[b]{0.3\textwidth}
   \centering
   \input{img/50Mbps/avg_delay.tex}
   \caption{Average delay ($R=50$~Mbps)}
   \label{fig:50Mbps_avg_delay}
  \end{subfigure}
  \hspace*{\fill}%
  \begin{subfigure}[b]{0.3\textwidth}
   \centering
   \input{img/100Mbps/avg_delay.tex}
   \caption{Average delay ($R=100$~Mbps)}
   \label{fig:100Mbps_avg_delay}
  \end{subfigure}
  \hspace*{\fill}%
  \begin{subfigure}[b]{0.3\textwidth}
   \centering
   \input{img/200Mbps/avg_delay.tex}
   \caption{Average delay ($R=200$~Mbps)}
   \label{fig:200Mbps_avg_delay}
  \end{subfigure}
  \hspace*{\fill}%
 
  \caption{Performance of the four scheduling configurations using a bursty application with period equal to $T=102.4$~ms, and an offered rate $R$ for each user.}
  \label{fig:fixed_app_rate}
 \end{figure*}

Finally, for \gls{sp} configuration \#3, where \textit{smart start} is not used but \glspl{sta} with scheduled \glspl{sp} are allowed to also access the \gls{cbap}, the performance is lower bounded by the CBAP-only scheduling and upper bounded by SP configuration \#2.
In fact, application bursts can either start during an ongoing SP or a CBAP and thus have to be split among different SPs or CBAPs.
For low traffic loads $\eta$, traffic will mostly be sent during the CBAP, mimicking the CBAP-only scheduler's performance.
Instead, considering node $k$, as $\eta$ increases, SPs allocated for nodes $\neq k$ will prevent it from freely transmitting over the whole BI, forcing it to either wait for its next SP or to concur with an increasingly busy and shorter CBAP, getting closer to the behavior (and the performance) of SP configuration \#2.
Contrary to what happens for the CBAP-only scheduler, though, the \gls{pcpap} has a way to control the traffic flow by rejecting ADDTS Requests, preventing the traffic from becoming unstable even for higher loads, at the cost of possibly denying some STAs to transmit.

In \cref{fig:burst_avg_delay_variation} we show the \textit{jitter} performance for the first scenario.
Again, as expected, CBAP-only scheduling shows an increasing jitter with an increasingly loaded network and becomes unstable for $\eta>0.8$, while SP configuration \#1 shows constant jitter irrespective of the traffic load, always lower than any other scheduling schemes.

Similar to what happened for the average delay, SP configuration \#2 has to account for two opposing trends.
Note that bursts starting during the CBAP will have extremely low jitter since they will be sent entirely during the next SP.
On one hand, a lower $\eta$ translates to shorter SPs, making it more likely for application bursts to start during a CBAP.
Bursts starting during a CBAP will be sent entirely during the next SP, resulting in a low jitter, while those starting during an SP will have to be split among two consecutive SPs, making one packet increase the jitter significantly.
On the other hand, a lower $\eta$ also reduces the burst size and, conversely, the number of packets composing the burst, making the single packet with higher delay variation weigh more in the average and thus affecting the jitter.
This second effect appears to be predominant and thus the jitter decreases as the traffic load increases.

SP configuration \#3 shows higher jitter than CBAP-only for lower values of $\eta$, since other nodes' SPs possibly interfere with the transmission of a full uninterrupted burst, while higher values of $\eta$ show a decreasing jitter.
This suggests that as the CBAP is reduced to leave space for the allocated SPs, nodes will be forced to use it less in favor of their allocated SPs, where transmissions are ensured and more stable but at the cost of a higher delay.

Finally, we show the \textit{aggregated throughput} normalized by the offered traffic in \cref{fig:burst_norm_thr}.
Clearly, all SP configurations can fully allocate the BI, resulting in unit normalized throughput.
The only exception to this is SP configuration \#3: allowing allocated users to also exploit the CBAP resources might prevent new users from transmitting in a timely fashion.
In fact, for high traffic loads, not only is the CBAP greatly reduced, but allocated STAs also contend for those resources, starving new users who might want to transmit non-QoS traffic or, as it happens in this case, send an ADDTS Request to schedule additional SPs, an event that clearly cannot happen when allocated users do not exploit CBAP resources.

Instead, the CBAP-only scheduler can only withstand the traffic demand for $\eta \leq 0.8$, then, as also noted for other metrics, the \gls{wifi} contention mechanism loses its effectiveness making the traffic unstable and starting to lose packets.


\paragraph{Second scenario} 
\label{par:second_scenario}

In \cref{fig:fixed_app_rate} we show the performance for the second scenario, where a fixed application rate was considered with a varying number of users.

Clearly, since MCS~4 was used with a PHY rate of $1\,155$~Mbps, for rates $R=50$~and $100$~Mbps, all scheduled SP allocations were able to meet the offered data rate (see \cref{fig:50Mbps_aggr_thr,fig:100Mbps_aggr_thr}).
Only SP configuration \#3 was not fully able to support the full 1~Gpbs as previously discussed for the first scenario.
Furthermore, also the CBAP-only case was unable to meet the aggregate demand since 1~Gbps of offered traffic or more corresponds to $\eta>0.8$ and, as suggested by the results shown for the first scenario, is thus unstable.

Regarding the average delay performance shown in \cref{fig:50Mbps_avg_delay,fig:100Mbps_avg_delay,fig:200Mbps_avg_delay}, similar results to the first scenario can be observed.

Using only the CBAP yields good performance for low traffic loads, which in this case corresponds to a lower number of users, while it remains unstable for high traffic loads.

Instead, while SP configuration \#1 is the lower bound achievable by any configuration consistently across all cases, SP configuration \#2 is the upper bound for all SP configurations.
As the offered traffic load increases, i.e., as more STAs transmit with higher application rates $R$, SP configuration \#3 tends to have the same performance as SP configuration \#2, as less CBAP is available.



\paragraph{Third scenario} 
\label{par:third_scenario}

\begin{figure}[t!]
 \newcommand\fheight{0.5\columnwidth}
 \newcommand\fwidth{1\columnwidth}
   \centering
   \input{img/burst_stdev/avg_delay.tex}
   \caption{Average delay of the different scheduling configurations with a bursty application with normally distributed period, with mean equal to $T=102.4$~ms and standard deviation equal to a fraction of its mean $\sigma=\rho T$.}
   \label{fig:burst_stdev_avg_delay}
 \end{figure}
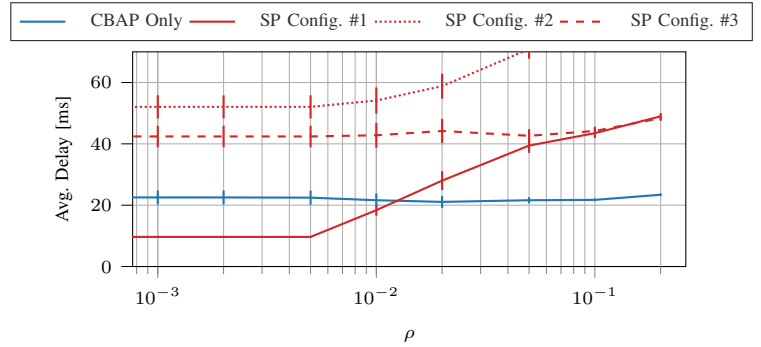

\cref{fig:burst_stdev_avg_delay} shows the average delay for the third proposed scenario, where we compare the four scheduling strategies considering a load of $\eta=0.75$ and an application with random period against its \textit{period deviation ratio} $\rho$, as described in \cref{sec:simulation_setup}.
Note that $\rho=0$ coincides with a deterministic application.

As expected, the CBAP-only case is not affected by the random periodicity of the application.

Similar to the first scenario, SP configuration \#3 shows worse performance than the CBAP-only scheduler, as users can only transmit in their own SPs or during the CBAP.
Since the applications are not synchronized with the SPs to begin with, also in this case the performance is not affected by the random periodicity of the application.

On the other hand, SP configuration \#1, appears to be optimal only for almost-deterministic applications, i.e., for extremely low values of $\rho$.
In fact, \textit{smart start} only synchronizes the application with the first allocated SP, meaning that if an application has a random period, bursts starting from the second one will be out of sync.
Since we allowed STAs to use the CBAP for SP configuration \#1, as $\rho$ increases, performance gets worse reaching the same average delay as SP configuration \#3, where cross-layer alignment is not enabled.

Even worse, SP configuration \#2 shows by far the worst behavior.
Not only is its performance bad for the deterministic case, but since STAs are only allowed to transmit during their own SPs and the SP duration was computed to be exactly the time required to send the whole burst, the random periodicity of the application further worsens the performance. 
In fact, if one period is longer than $T$, part of an SP might never be used, although the average traffic will still require all SPs to be fully utilized.
The more random the application, the more likely this event, possibly leaving more and more portions of SPs not utilized.



\paragraph{Results Overview} 
\label{par:results_overview}

To summarize, the simulation results show that when the network load is low, contention-based channel access is capable of yielding overall good performance, but as the amount of offered traffic increases, average delay and jitter are quickly affected.
Instead, SP scheduling shows its full potential only when cross-layer information is exchanged between the APP and the MAC layer, allowing the application to synchronize with the scheduled SPs.

Furthermore, we showed that if (i) the application and the SP allocations share the same period $T$, (ii) STAs can only transmit during their own SPs, (iii) the SP duration coincides with the time required to transmit the burst, and (iv) the application start time is uniformly distributed over a period $T$, then the average delay is equal to $T/2$, irrespective of the burst size, the number of users, or the network traffic load.

On top of this, SP scheduling allows the \gls{pcpap} to accept and reject incoming traffic flows, allowing better control of the network even in the most intensive traffic regimes, thus being able to ensure to a limited number of users the required amount of resources without making the transmission unstable, unlike contention-based access alone.

Finally, we showed that small amounts of randomness in the period duration can easily favor the simpler contention-based access over the more complex SP scheduling, but further studies need to be done as the setup was extremely simple.



\section{Conclusions} 
\label{sec:conclusions}

In this paper, we presented an open-source scheduling framework for \gls{wigig} based on the ns-3 implementation of the IEEE~802.11ad standard~\cite{ns3-802.11ad-scheduling}.
We implemented two schedulers, one based on contention-based channel access, the other based on periodic SP allocations, and compared their performance on three different scenarios. 
Results show that SP scheduling is able to surpass contention-based channel access and yield the best performance only when cross-layer information between the MAC and APP layers is exchanged. 
Moreover, adding even small amounts of randomness to the periodic application results in great performance degradation for periodic SP scheduling, making contention based-access the preferred option in most cases.
Future works will mainly focus on more complex scenarios and schedulers involving, e.g., more realistic traffic models or dynamic \gls{mmw} channel modeling.


\bibliographystyle{IEEEtran}
\bibliography{bibl.bib}

\end{document}

%% file: myTikz.tex
\pgfplotsset{compat=newest}
\pgfplotsset{plot coordinates/math parser=false}
\pgfplotsset{every axis/.append style={
                    label style={font=\scriptsize},
                    tick label style={font=\scriptsize},
                    legend style={font=\scriptsize}
                    }}
\usetikzlibrary{plotmarks,shapes,patterns,decorations.pathreplacing,backgrounds,calc,arrows,arrows.meta,spy,matrix,shadows,trees,positioning}
\usepgfplotslibrary{patchplots,groupplots}

\tikzstyle{startstop} = [rectangle, rounded corners, minimum width=2cm, minimum height=0.5cm,text centered, draw=black]
\tikzstyle{io} = [trapezium, trapezium left angle=70, trapezium right angle=110, minimum width=3cm, minimum height=1cm, text centered, draw=black]
\tikzstyle{process} = [rectangle, minimum width=2cm, minimum height=0.5cm, text centered, draw=black, alignb=center]
\tikzstyle{decision} = [ellipse, minimum width=2cm, minimum height=1cm, text centered, draw=black]
\tikzstyle{arrow} = [thick,<->,>=stealth]
\tikzstyle{line} = [thick,>=stealth]
\tikzstyle{darrow} = [thick,<->,>=stealth,dashed]
\tikzstyle{sarrow} = [thick,->,>=stealth]
\tikzstyle{larrow} = [line width=0.1mm,dashdotted,->,>=stealth]

\makeatletter
\def\grd@save@target#1{%
  \def\grd@target{#1}}
\def\grd@save@start#1{%
  \def\grd@start{#1}}
\tikzset{
  grid with coordinates/.style={
    to path={%
      \pgfextra{%
        \edef\grd@@target{(\tikztotarget)}%
        \tikz@scan@one@point\grd@save@target\grd@@target\relax
        \edef\grd@@start{(\tikztostart)}%
        \tikz@scan@one@point\grd@save@start\grd@@start\relax
        \draw[minor help lines] (\tikztostart) grid (\tikztotarget);
        \draw[major help lines] (\tikztostart) grid (\tikztotarget);
        \grd@start
        \pgfmathsetmacro{\grd@xa}{\the\pgf@x/1cm}
        \pgfmathsetmacro{\grd@ya}{\the\pgf@y/1cm}
        \grd@target
        \pgfmathsetmacro{\grd@xb}{\the\pgf@x/1cm}
        \pgfmathsetmacro{\grd@yb}{\the\pgf@y/1cm}
        \pgfmathsetmacro{\grd@xc}{\grd@xa + \pgfkeysvalueof{/tikz/grid with coordinates/major step x}}
        \pgfmathsetmacro{\grd@yc}{\grd@ya + \pgfkeysvalueof{/tikz/grid with coordinates/major step y}}
        \foreach \x in {\grd@xa,\grd@xc,...,\grd@xb}
        \node[anchor=north] at (\x,\grd@ya) {\pgfmathprintnumber{\x}};
        \foreach \y in {\grd@ya,\grd@yc,...,\grd@yb}
        \node[anchor=east] at (\grd@xa,\y) {\pgfmathprintnumber{\y}};
      }
    }
  },
  minor help lines/.style={
    help lines,
    gray,
    line cap =round,
    xstep=\pgfkeysvalueof{/tikz/grid with coordinates/minor step x},
    ystep=\pgfkeysvalueof{/tikz/grid with coordinates/minor step y}
  },
  major help lines/.style={
    help lines,
    line cap =round,
    line width=\pgfkeysvalueof{/tikz/grid with coordinates/major line width},
    xstep=\pgfkeysvalueof{/tikz/grid with coordinates/major step x},
    ystep=\pgfkeysvalueof{/tikz/grid with coordinates/major step y}
  },
  grid with coordinates/.cd,
  minor step x/.initial=.5,
  minor step y/.initial=.2,
  major step x/.initial=1,
  major step y/.initial=1,
  major line width/.initial=1pt,
}
\makeatother

%% file: acronyms.tex
\newacronym{3gpp}{3GPP}{3rd Generation Partnership Project}
\newacronym{5g}{5G}{5\textsuperscript{th} Generation}
\newacronym{5gc}{5GC}{5G Core}
\newacronym{bs}{BS}{Base Station}
\newacronym{abft}{A-BFT}{Association-BeamForming Training}
\newacronym[firstplural=Access Categories (ACs)]{ac}{AC}{Access Category}
\newacronym{adc}{ADC}{Analog to Digital Converter}
\newacronym{addts}{ADDTS}{Add Traffic Stream}
\newacronym{afbw}{AFBW}{Average Fading Bandwidth}
\newacronym{aid}{AID}{Association ID}
\newacronym{aimd}{AIMD}{Additive Increase Multiplicative Decrease}
\newacronym{am}{AM}{Acknowledged Mode}
\newacronym{amc}{AMC}{Adaptive Modulation and Coding}
\newacronym{ampdu}{A-MPDU}{MAC Protocol Data Unit Aggregation}
\newacronym{aoa}{AoA}{Angle of Arrival}
\newacronym{aod}{AoD}{Angle of Departure}
\newacronym{ap}{AP}{Access Point}
\newacronym{app}{APP}{Application Layer}
\newacronym{aqm}{AQM}{Active Queue Management}
\newacronym{ar}{AR}{Augmented Reality}
\newacronym{arf}{ARF}{Auto Rate Fallback}
\newacronym{arp}{ARP}{Address Resolution Protocol}
\newacronym{ati}{ATI}{Announcement Transmission Interval}
\newacronym{awgn}{AGWN}{Additive White Gaussian Noise}
\newacronym{awv}{AWV}{Antenna Weight Vector}
\newacronym{balia}{BALIA}{Balanced Link Adaptation}
\newacronym{bdp}{BDP}{Bandwidth-Delay Product}
\newacronym{ber}{BER}{Bit Error Rate}
\newacronym{bf}{BF}{Beamforming}
\newacronym{bframe}{B-frame}{Bipredictive-coded frame}
\newacronym{bhi}{BHI}{Beacon Header Interval}
\newacronym{bi}{BI}{Beacon Interval}
\newacronym{brp}{BRP}{Beam Refinement Protocol}
\newacronym{bss}{BSS}{Basic Service Set}
\newacronym{bti}{BTI}{Beacon Transmission Interval}
\newacronym{cad}{CAD}{Computer-aided Design}
\newacronym{cbap}{CBAP}{Contention-Based Access Period}
\newacronym{cbr}{CBR}{Constant Bitrate}
\newacronym{cc}{CC}{Congestion Control}
\newacronym{cdf}{CDF}{Cumulative Distribution Function}
\newacronym{cir}{CIR}{Channel Impulse Response}
\newacronym{cn}{CN}{Core Network}
\newacronym{cp}{CP}{Control Plane}
\newacronym{cqi}{CQI}{Channel Quality Indicator}
\newacronym{crs}{CRS}{Cell Reference Signal}
\newacronym{csirs}{CSI-RS}{Channel State Information - Reference Signal}
\newacronym{csmaca}{CSMA/CA}{Carrier Sense Multiple Access with Collision Avoidance}
\newacronym{cts}{CTS}{Clear to Send}
\newacronym{dc}{DC}{Dual Connectivity}
\newacronym{dce}{DCE}{Direct Code Execution}
\newacronym{dcf}{DCF}{Distributed Coordination Function}
\newacronym{dci}{DCI}{Downlink Control Information}
\newacronym{delts}{DELTS}{Delete Traffic Stream}
\newacronym{dl}{DL}{Downlink}
\newacronym{dmg}{DMG}{Directional Multi-Gigabit}
\newacronym{dmr}{DMR}{Deadline Miss Ratio}
\newacronym{dmrs}{DMRS}{DeModulation Reference Signal}
\newacronym{dti}{DTI}{Data Transmission Interval}
\newacronym{e2e}{E2E}{End-to-End}
\newacronym{ecn}{ECN}{Explicit Congestion Notification}
\newacronym{edca}{EDCA}{Enhanced Distributed Channel Access}
\newacronym{edf}{EDF}{Earliest Deadline First}
\newacronym{embb}{eMBB}{Enhanced Mobile BroadBand}
\newacronym{enb}{eNB}{evolved Node Base}
\newacronym{endc}{EN-DC}{E-UTRAN-\gls{nr} \gls{dc}}
\newacronym{epc}{EPC}{Evolved Packet Core}
\newacronym{es}{ES}{Edge Server}
\newacronym{ese}{ESE}{Extended Schedule Element}
\newacronym{fdd}{FDD}{Frequency Division Duplexing}
\newacronym{fdma}{FDMA}{Frequency Division Multiple Access}
\newacronym{fec}{FEC}{Forward Error Correction}
\newacronym{fov}{FoV}{Field-of-View}
\newacronym{fs}{FS}{Fast Switching}
\newacronym{ftp}{FTP}{File Transfer Protocol}
\newacronym{fr2}{FR2}{Frequency Range 2}
\newacronym{gmm}{GMM}{Gaussian Mixture Model}
\newacronym{gnb}{gNB}{Next Generation Node Base}
\newacronym[firstplural=Group of Pictures (GoPs)]{gop}{GoP}{Group of Pictures}
\newacronym{harq}{HARQ}{Hybrid Automatic Repeat reQuest}
\newacronym{hetnet}{HetNet}{Heterogeneous Network}
\newacronym{hh}{HH}{Hard Handover}
\newacronym{hol}{HOL}{Head-of-Line}
\newacronym{hqf}{HQF}{Highest-quality-first}
\newacronym{ia}{IA}{Initial Access}
\newacronym{iab}{IAB}{Integrated Access and Backhaul}
\newacronym{ibss}{IBSS}{Independent Basic Service Set}
\newacronym{id}{ID}{Identifier}
\newacronym{ifi}{IFI}{Inter-Frame Inter-arrival}
\newacronym{iframe}{I-frame}{Intra-coded frame}
\newacronym{imt}{IMT}{International Mobile Telecommunication}
\newacronym{imt2020}{IMT-2020}{International Mobile Te\-le\-com\-mu\-ni\-ca\-tion-2020}
\newacronym{inr}{INR}{Interference to Noise Ratio}
\newacronym{iot}{IoT}{Internet of Things}
\newacronym{ipa}{IPA}{Inter-Packet Arrival}
\newacronym{ism}{ISM}{Industrial, Scientific, and Medical}
\newacronym{itu}{ITU}{International Telecommunication Union}
\newacronym{kpi}{KPI}{Key Performance Indicator}
\newacronym{lcf}{LCF}{Level Crossing Frequency}
\newacronym{lcm}{lcm}{least common multiple}
\newacronym{lcr}{LCR}{Level Crossing Rate}
\newacronym{los}{LoS}{Line-of-Sight}
\newacronym{lp}{LP}{Low Power}
\newacronym{lte}{LTE}{Long Term Evolution}
\newacronym{m2m}{M2M}{Machine to Machine}
\newacronym{mac}{MAC}{Medium Access Control}
\newacronym{mc}{MC}{Multi-Connectivity}
\newacronym{mcs}{MCS}{Modulation and Coding Scheme}
\newacronym{mec}{MEC}{Mobile Edge Cloud}
\newacronym{mi}{MI}{Mutual Information}
\newacronym{mib}{MIB}{Master Information Block}
\newacronym{mimo}{MIMO}{Multiple Input, Multiple Output}
\newacronym{mumimo}{MU-MIMO}{Multi-User Multiple Input, Multiple Output}
\newacronym{ml}{ML}{Machine Learning}
\newacronym{mlr}{MLR}{Maximum-local-rate}
\newacronym[plural=\gls{mme}s,firstplural=Mobility Management Entities (MMEs)]{mme}{MME}{Mobility Management Entity}
\newacronym{mmw}{mmW}{Millimeter Wave}
\newacronym{moi}{MoI}{Method of Images}
\newacronym{mpc}{MPC}{Multi Path Component}
\newacronym{mpdu}{MPDU}{MAC Protocol Data Unit}
\newacronym{mptcp}{MPTCP}{Multipath TCP}
\newacronym{mr}{MR}{Maximum Rate}
\newacronym{mrdc}{MR-DC}{Multi \gls{rat} \gls{dc}}
\newacronym{msdu}{MSDU}{MAC Service Data Unit}
\newacronym{mss}{MSS}{Maximum Segment Size}
\newacronym{mtd}{MTD}{Machine-Type Device}
\newacronym{mtu}{MTU}{Maximum Transmission Unit}
\newacronym{nav}{NAV}{Network Allocation Vector}
\newacronym{ncbr}{NCBR}{Non-Constant Bitrate}
\newacronym{nfv}{NFV}{Network Function Virtualization}
\newacronym{nlos}{NLoS}{Non-Line-of-Sight}
\newacronym{nr}{NR}{New Radio}
\newacronym{nrmse}{NRMSE}{Normalized Root Mean Square Error}
\newacronym{ns2}{ns-2}{Network Simulator 2}
\newacronym{ns3}{ns-3}{Network Simulator 3}
\newacronym{nsa}{NSA}{Non Stand Alone}
\newacronym{o2i}{O2I}{Outdoor-to-Indoor}
\newacronym{ofdm}{OFDM}{Orthogonal Frequency Division Multiplexing}
\newacronym{pa}{PA}{Position-aware}
\newacronym{pan}{PAN}{Personal Area Network}
\newacronym{pbch}{PBCH}{Physical Broadcast Channel}
\newacronym{pbss}{PBSS}{Personal Basic Service Set}
\newacronym{pcf}{PCF}{Point Coordinator Function}
\newacronym{pcp}{PCP}{\gls{pbss} Central Point}
\newacronym{pcpap}{PCP/AP}{\acrlong{pcp}/\acrlong{ap}}
\newacronym{pdcch}{PDCCH}{Physical Downlonk Control Channel}
\newacronym{pdcp}{PDCP}{Packet Data Convergence Protocol}
\newacronym{pdf}{PDF}{Probability Distribution Function}
\newacronym{pdsch}{PDSCH}{Physical Downlink Shared Channel}
\newacronym{per}{PER}{Packet Error Rate}
\newacronym{pdu}{PDU}{Packet Data Unit}
\newacronym{pf}{PF}{Proportional Fair}
\newacronym{pframe}{P-frame}{Predictive-coded frame}
\newacronym{pgw}{PGW}{Packet Gateway}
\newacronym{phy}{PHY}{Physical Layer}
\newacronym{ppdu}{PPDU}{PHY Protocol Data Unit}
\newacronym{ppp}{PPP}{Poisson Point Process}
\newacronym{prb}{PRB}{Physical Resource Block}
\newacronym{pss}{PSS}{Primary Synchronization Signal}
\newacronym{pucch}{PUCCH}{Physical Uplink Control Channel}
\newacronym{pusch}{PUSCH}{Physical Uplink Shared Channel}
\newacronym{qd}{QD}{Quasi Deterministic}
\newacronym{qoe}{QoE}{Quality of Experience}
\newacronym{qos}{QoS}{Quality of Service}
\newacronym{rach}{RACH}{Random Access Channel}
\newacronym{ran}{RAN}{Radio Access Network}
\newacronym[firstplural=Radio Access Technologies (RATs)]{rat}{RAT}{Radio Access Technology}
\newacronym{red}{RED}{Random Early Detection}
\newacronym{rf}{RF}{Radio Frequency}
\newacronym{rl}{RL}{Reinforcement Learning}
\newacronym{rlc}{RLC}{Radio Link Control}
\newacronym{rlf}{RLF}{Radio Link Failure}
\newacronym{rr}{RR}{Round Robin}
\newacronym{rrc}{RRC}{Radio Resource Control}
\newacronym{rrm}{RRM}{Radio Resource Management}
\newacronym{rs}{RS}{Remote Server}
\newacronym{rsrp}{RSRP}{Reference Signal Received Power}
\newacronym{rsrq}{RSRQ}{Reference Signal Received Quality}
\newacronym{rss}{RSS}{Received Signal Strength}
\newacronym{rssi}{RSSI}{Received Signal Strength Indicator}
\newacronym{rt}{RT}{Ray Tracer}
\newacronym{rts}{RTS}{Request to Send}
\newacronym{rtt}{RTT}{Round Trip Time}
\newacronym{rw}{RW}{Receive Window}
\newacronym{rx}{RX}{Receiver}
\newacronym{sa}{SA}{standalone}
\newacronym{sack}{SACK}{Selective Acknowledgment}
\newacronym{sap}{SAP}{Service Access Point}
\newacronym{sc}{SC}{Single Carrier}
\newacronym{sch}{SCH}{Secondary Cell Handover}
\newacronym{scm}{SCM}{Spatial Channel Model}
\newacronym{scoot}{SCOOT}{Split Cycle Offset Optimization Technique}
\newacronym{sdma}{SDMA}{Spatial Division Multiple Access}
\newacronym{sdr}{SDR}{Software Defined Radio}
\newacronym{semm}{SEMM}{SPCA-EDCA Mixed Mode}
\newacronym{si}{SI}{Study Item}
\newacronym{sib}{SIB}{Secondary Information Block}
\newacronym{sinr}{SINR}{Signal-to-Interference-plus-Noise Ratio}
\newacronym{sir}{SIR}{Signal-to-Interference Ratio}
\newacronym{sls}{SLS}{Sector-Level Sweep}
\newacronym{sm}{SM}{Saturation Mode}
\newacronym{snr}{SNR}{Signal-to-Noise Ratio}
\newacronym{son}{SON}{Self-Organizing Network}
\newacronym{sp}{SP}{Service Period}
\newacronym{spr}{SPR}{Service Period Request}
\newacronym{srs}{SRS}{Sounding Reference Signal}
\newacronym{ss}{SS}{Synchronization Signal}
\newacronym{sss}{SSS}{Secondary Synchronization Signal}
\newacronym{ssw}{SSW}{Sector Sweep}
\newacronym{sta}{STA}{Station}
\newacronym{stb}{STB}{Set Top Box}
\newacronym{tb}{TB}{Transport Block}
\newacronym[firstplural=Traffic Categories (TCs)]{tc}{TC}{Traffic Category}
\newacronym{tbtt}{TBTT}{Target Beacon Transmission Time}
\newacronym{tcp}{TCP}{Transmission Control Protocol}
\newacronym{tdd}{TDD}{Time Division Duplexing}
\newacronym{tdma}{TDMA}{Time Division Multiple Access}
\newacronym{tfl}{TfL}{Transport for London}
\newacronym{tgad}{TGad}{Task Group ad}
\newacronym{tgay}{TGay}{Task Group ay}
\newacronym{tsconst}{TSCONST}{Traffic Scheduling Constraint}
\newacronym{tsf}{TSF}{Timing Synchronization Function}
\newacronym{tm}{TM}{Transparent Mode}
\newacronym{trp}{TRP}{Transmitter Receiver Pair}
\newacronym{ts}{TS}{Traffic Stream}
\newacronym{tspec}{TSPEC}{Traffic Specification}
\newacronym{tti}{TTI}{Transmission Time Interval}
\newacronym{ttt}{TTT}{Time-to-Trigger}
\newacronym{tx}{TX}{Transmitter}
\newacronym[firstplural=Transmission Opportunities (TXOPs)]{txop}{TXOP}{Transmission Opportunity}
\newacronym{udp}{UDP}{User Datagram Protocol}
\newacronym{ue}{UE}{User Equipment}
\newacronym{ul}{UL}{Uplink}
\newacronym{um}{UM}{Unacknowledged Mode}
\newacronym{uma}{UMa}{Urban Macro}
\newacronym{uml}{UML}{Unified Modeling Language}
\newacronym{up}{UP}{User Priority}
\newacronym{utc}{UTC}{Urban Traffic Control}
\newacronym{vbr}{VBR}{Variable Bit Rate}
\newacronym{vm}{VM}{Virtual Machine}
\newacronym{vr}{VR}{Virtual Reality}
\newacronym{wbf}{WBF}{Wired Bias Function}
\newacronym{wf}{WF}{Wired-first}
\newacronym{wifi}{Wi-Fi}{Wireless Fidelity}
\newacronym{wigig}{WiGig}{Wireless Gigabit}
\newacronym{wlan}{WLAN}{Wireless Local Area Network}
\newacronym{xr}{XR}{eXtended Reality}
\newacronym{aifs}{AIFS}{Arbitration Inter-Frame Space}

%% file: img/legend.tex
%
%
\definecolor{color0}{rgb}{0.12156862745098,0.466666666666667,0.705882352941177}
\definecolor{color1}{rgb}{0.83921568627451,0.152941176470588,0.156862745098039}
\begin{tikzpicture}
\pgfplotsset{every tick label/.append style={font=\scriptsize}}

\begin{axis}[%
width=0,
height=0,
at={(0,0)},
scale only axis,
xmin=0,
xmax=0,
xtick={},
ymin=0,
ymax=0,
ytick={},
axis background/.style={fill=white},
legend style={legend cell align=center,
              align=center,
              draw=white!15!black,
              at={(0, 0)},
              anchor=center,
              /tikz/every even column/.append style={column sep=2em}},
legend columns=4,
]

\addplot [thick, color0]
table {%
0 0
};
\addlegendentry{CBAP Only}
\addplot [thick, color1]
table {%
0 0
};
\addlegendentry{SP Config. \#1}
\addplot [thick, color1, densely dotted]
table {%
0 0
};
\addlegendentry{SP Config. \#2}
\addplot [thick, color1, dashed]
table {%
0 0
};
\addlegendentry{SP Config. \#3}

\end{axis}
\end{tikzpicture}%

%% file: img/burst/avg_delay.tex
\begin{tikzpicture}

\definecolor{color0}{rgb}{0.12156862745098,0.466666666666667,0.705882352941177}
\definecolor{color1}{rgb}{0.83921568627451,0.152941176470588,0.156862745098039}

\begin{axis}[
width=\fwidth,
height=\fheight,
legend cell align={left},
legend style={fill opacity=0.8, draw opacity=1, text opacity=1, at={(0.03,0.97)}, anchor=north west, draw=white!80!black},
tick align=outside,
tick pos=left,
x grid style={white!69.0196078431373!black},
xlabel={$\eta$},
xmajorgrids,
xmin=-0.0395, xmax=1.0495,
xtick style={color=black},
xtick={-0.2,0,0.2,0.4,0.6,0.8,1,1.2},
xticklabels={−0.2,0.0,0.2,0.4,0.6,0.8,1.0,1.2},
y grid style={white!69.0196078431373!black},
ylabel={Avg. Delay [ms]},
ymajorgrids,
ymin=-5, ymax=60,
ytick style={color=black}
]
\path [draw=color0, thick]
(axis cs:0.01,0.205787208870789)
--(axis cs:0.01,0.391520807707499);

\path [draw=color0, thick]
(axis cs:0.1,1.57786407235494)
--(axis cs:0.1,2.03471656632943);

\path [draw=color0, thick]
(axis cs:0.2,3.30848058282302)
--(axis cs:0.2,4.07059940399268);

\path [draw=color0, thick]
(axis cs:0.3,5.46448140308823)
--(axis cs:0.3,6.76553100936564);

\path [draw=color0, thick]
(axis cs:0.4,7.84255459582289)
--(axis cs:0.4,9.77614569108658);

\path [draw=color0, thick]
(axis cs:0.5,10.6259983425496)
--(axis cs:0.5,13.2889811363036);

\path [draw=color0, thick]
(axis cs:0.6,13.8887945990804)
--(axis cs:0.6,17.3798657641615);

\path [draw=color0, thick]
(axis cs:0.7,17.871181202477)
--(axis cs:0.7,22.1126047733454);

\path [draw=color0, thick]
(axis cs:0.8,22.7458761066876)
--(axis cs:0.8,27.638299391215);

\path [draw=color0, thick]
(axis cs:0.9,194.737660512901)
--(axis cs:0.9,230.207469040384);


\path [draw=color1, thick]
(axis cs:0.01,0.191357277619612)
--(axis cs:0.01,0.20045396372514);

\path [draw=color1, thick]
(axis cs:0.1,1.30947735482451)
--(axis cs:0.1,1.31043147142108);

\path [draw=color1, thick]
(axis cs:0.2,2.59152829978824)
--(axis cs:0.2,2.59200740199628);

\path [draw=color1, thick]
(axis cs:0.3,3.87527899908199)
--(axis cs:0.3,3.87559946306518);

\path [draw=color1, thick]
(axis cs:0.4,5.15947288344969)
--(axis cs:0.4,5.15971397381558);

\path [draw=color1, thick]
(axis cs:0.5,6.44356317314064)
--(axis cs:0.5,6.44375881844441);

\path [draw=color1, thick]
(axis cs:0.6,7.7228102813797)
--(axis cs:0.6,7.72297207308311);

\path [draw=color1, thick]
(axis cs:0.7,9.00712985052406)
--(axis cs:0.7,9.00727265649652);

\path [draw=color1, thick]
(axis cs:0.8,10.2915583092621)
--(axis cs:0.8,10.2916803408592);

\path [draw=color1, thick]
(axis cs:0.9,11.5759889874662)
--(axis cs:0.9,11.5760995112733);


\path [draw=color1, thick]
(axis cs:0.01,0.352563936957852)
--(axis cs:0.01,0.741897064739545);

\path [draw=color1, thick]
(axis cs:0.1,1.99184063295332)
--(axis cs:0.1,2.77999670831234);

\path [draw=color1, thick]
(axis cs:0.2,4.90270457051287)
--(axis cs:0.2,6.41771308511481);

\path [draw=color1, thick]
(axis cs:0.3,9.27560101862216)
--(axis cs:0.3,11.6018195633076);

\path [draw=color1, thick]
(axis cs:0.4,14.1530327684927)
--(axis cs:0.4,17.1257871906197);

\path [draw=color1, thick]
(axis cs:0.5,19.9517719380955)
--(axis cs:0.5,24.2982543688014);

\path [draw=color1, thick]
(axis cs:0.6,26.7764769175532)
--(axis cs:0.6,33.3739441258951);

\path [draw=color1, thick]
(axis cs:0.7,35.0405711975811)
--(axis cs:0.7,41.5648636439923);

\path [draw=color1, thick]
(axis cs:0.8,42.2611585121621)
--(axis cs:0.8,49.7074440418743);

\path [draw=color1, thick]
(axis cs:0.9,47.3631388822112)
--(axis cs:0.9,55.918137277617);


\path [draw=color1, thick]
(axis cs:0.01,44.6837163153675)
--(axis cs:0.01,55.8699725052268);

\path [draw=color1, thick]
(axis cs:0.1,47.0905517946929)
--(axis cs:0.1,58.4924789127217);

\path [draw=color1, thick]
(axis cs:0.2,47.4337027205709)
--(axis cs:0.2,59.3977917048263);

\path [draw=color1, thick]
(axis cs:0.3,47.0771592279094)
--(axis cs:0.3,56.9701292438626);

\path [draw=color1, thick]
(axis cs:0.4,46.4598262281854)
--(axis cs:0.4,56.4552274076662);

\path [draw=color1, thick]
(axis cs:0.5,46.8084715297775)
--(axis cs:0.5,55.7737529843184);

\path [draw=color1, thick]
(axis cs:0.6,46.9632846384876)
--(axis cs:0.6,55.0244344598635);

\path [draw=color1, thick]
(axis cs:0.7,47.6130782594789)
--(axis cs:0.7,54.9317950286115);

\path [draw=color1, thick]
(axis cs:0.8,48.6207389657365)
--(axis cs:0.8,56.3884516566091);

\path [draw=color1, thick]
(axis cs:0.9,49.5482558313366)
--(axis cs:0.9,57.702399521211);


\addplot [thick, color0]
table {%
0.01 0.298654008289144
0.1 1.80629031934219
0.2 3.68953999340785
0.3 6.11500620622693
0.4 8.80935014345474
0.5 11.9574897394266
0.6 15.6343301816209
0.7 19.9918929879112
0.8 25.1920877489513
0.9 212.472564776643
};
\addplot [thick, color1]
table {%
0.01 0.195905620672376
0.1 1.3099544131228
0.2 2.59176785089226
0.3 3.87543923107359
0.4 5.15959342863264
0.5 6.44366099579253
0.6 7.72289117723141
0.7 9.00720125351029
0.8 10.2916193250607
0.9 11.5760442493698
};
\addplot [thick, color1, dashed]
table {%
0.01 0.547230500848699
0.1 2.38591867063283
0.2 5.66020882781384
0.3 10.4387102909649
0.4 15.6394099795562
0.5 22.1250131534484
0.6 30.0752105217242
0.7 38.3027174207867
0.8 45.9843012770182
0.9 51.6406380799141
};
\addplot [thick, color1, densely dotted]
table {%
0.01 50.2768444102971
0.1 52.7915153537073
0.2 53.4157472126986
0.3 52.023644235886
0.4 51.4575268179258
0.5 51.291112257048
0.6 50.9938595491755
0.7 51.2724366440452
0.8 52.5045953111728
0.9 53.6253276762738
};
\end{axis}

\end{tikzpicture}

%% file: img/burst/avg_delay_variation.tex
\begin{tikzpicture}

\definecolor{color0}{rgb}{0.12156862745098,0.466666666666667,0.705882352941177}
\definecolor{color1}{rgb}{0.83921568627451,0.152941176470588,0.156862745098039}

\begin{axis}[
width=\fwidth,
height=\fheight,
legend cell align={left},
legend style={fill opacity=0.8, draw opacity=1, text opacity=1, at={(0.03,0.97)}, anchor=north west, draw=white!80!black},
tick align=outside,
tick pos=left,
x grid style={white!69.0196078431373!black},
xlabel={$\eta$},
xmajorgrids,
xmin=-0.0395, xmax=1.0495,
xtick style={color=black},
xtick={-0.2,0,0.2,0.4,0.6,0.8,1,1.2},
xticklabels={−0.2,0.0,0.2,0.4,0.6,0.8,1.0,1.2},
y grid style={white!69.0196078431373!black},
ylabel={Jitter [$\mu$s]},
scaled y ticks={real:0.001}, 
ytick scale label code/.code={}, 
ymajorgrids,
ymin=0, ymax=0.2,
]
\path [draw=color0, thick]
(axis cs:0.01,0.0236193344048238)
--(axis cs:0.01,0.0360535214904381);

\path [draw=color0, thick]
(axis cs:0.1,0.0222704266302452)
--(axis cs:0.1,0.025756098284532);

\path [draw=color0, thick]
(axis cs:0.2,0.0224020662421334)
--(axis cs:0.2,0.0246467795551179);

\path [draw=color0, thick]
(axis cs:0.3,0.0262762431463635)
--(axis cs:0.3,0.0313768512061718);

\path [draw=color0, thick]
(axis cs:0.4,0.028111302633804)
--(axis cs:0.4,0.0345430002557978);

\path [draw=color0, thick]
(axis cs:0.5,0.0334100114168929)
--(axis cs:0.5,0.0416365215660018);

\path [draw=color0, thick]
(axis cs:0.6,0.0364773366479389)
--(axis cs:0.6,0.0468014868930361);

\path [draw=color0, thick]
(axis cs:0.7,0.0420576768046504)
--(axis cs:0.7,0.053456183721289);

\path [draw=color0, thick]
(axis cs:0.8,0.0550990101190962)
--(axis cs:0.8,0.0690847295210283);

\path [draw=color0, thick]
(axis cs:0.9,0.566974452037829)
--(axis cs:0.9,0.661109637576635);


\path [draw=color1, thick]
(axis cs:0.01,0.0165024303996992)
--(axis cs:0.01,0.0165937498116092);

\path [draw=color1, thick]
(axis cs:0.1,0.02048854157883)
--(axis cs:0.1,0.0204983594034481);

\path [draw=color1, thick]
(axis cs:0.2,0.0206947615898979)
--(axis cs:0.2,0.0206996816241085);

\path [draw=color1, thick]
(axis cs:0.3,0.0207862714286269)
--(axis cs:0.3,0.0207895557388694);

\path [draw=color1, thick]
(axis cs:0.4,0.0208150766598405)
--(axis cs:0.4,0.0208175425201886);

\path [draw=color1, thick]
(axis cs:0.5,0.0208467073342418)
--(axis cs:0.5,0.0208486914555765);

\path [draw=color1, thick]
(axis cs:0.6,0.0208690160592588)
--(axis cs:0.6,0.02087066424256);

\path [draw=color1, thick]
(axis cs:0.7,0.0208735980690464)
--(axis cs:0.7,0.0208750365566126);

\path [draw=color1, thick]
(axis cs:0.8,0.0208856567593145)
--(axis cs:0.8,0.0208868949439803);

\path [draw=color1, thick]
(axis cs:0.9,0.0208874168696196)
--(axis cs:0.9,0.0208885197497944);


\path [draw=color1, thick]
(axis cs:0.01,0.0312798327194502)
--(axis cs:0.01,0.0631829581579075);

\path [draw=color1, thick]
(axis cs:0.1,0.024569297922065)
--(axis cs:0.1,0.0305208354158978);

\path [draw=color1, thick]
(axis cs:0.2,0.0254670871502251)
--(axis cs:0.2,0.0317409588569727);

\path [draw=color1, thick]
(axis cs:0.3,0.0313133217274855)
--(axis cs:0.3,0.040273948726884);

\path [draw=color1, thick]
(axis cs:0.4,0.0365928195446286)
--(axis cs:0.4,0.0477995611268582);

\path [draw=color1, thick]
(axis cs:0.5,0.0385443121350768)
--(axis cs:0.5,0.0574164396322473);

\path [draw=color1, thick]
(axis cs:0.6,0.0477676677567447)
--(axis cs:0.6,0.0649911814889274);

\path [draw=color1, thick]
(axis cs:0.7,0.0498878819864318)
--(axis cs:0.7,0.0622071683270459);

\path [draw=color1, thick]
(axis cs:0.8,0.0398472099650279)
--(axis cs:0.8,0.049196145255374);

\path [draw=color1, thick]
(axis cs:0.9,0.0345613713499637)
--(axis cs:0.9,0.0417962164229479);


\path [draw=color1, thick]
(axis cs:0.01,0.989675373650617)
--(axis cs:0.01,1.2679453339571);

\path [draw=color1, thick]
(axis cs:0.1,0.117107577829871)
--(axis cs:0.1,0.156241548644884);

\path [draw=color1, thick]
(axis cs:0.2,0.0673077467993963)
--(axis cs:0.2,0.0955211536816443);

\path [draw=color1, thick]
(axis cs:0.3,0.0528346637926106)
--(axis cs:0.3,0.073982243775355);

\path [draw=color1, thick]
(axis cs:0.4,0.0452974578369055)
--(axis cs:0.4,0.0605825434664292);

\path [draw=color1, thick]
(axis cs:0.5,0.0414953343112935)
--(axis cs:0.5,0.0547760081788036);

\path [draw=color1, thick]
(axis cs:0.6,0.0369593548761599)
--(axis cs:0.6,0.0495369472015508);

\path [draw=color1, thick]
(axis cs:0.7,0.0334574160935851)
--(axis cs:0.7,0.0422461278159741);

\path [draw=color1, thick]
(axis cs:0.8,0.0327636264057788)
--(axis cs:0.8,0.0405796124665259);

\path [draw=color1, thick]
(axis cs:0.9,0.0328592123686329)
--(axis cs:0.9,0.0405381339960664);


\addplot [thick, color0]
table {%
0.01 0.0298364279476309
0.1 0.0240132624573886
0.2 0.0235244228986256
0.3 0.0288265471762676
0.4 0.0313271514448009
0.5 0.0375232664914474
0.6 0.0416394117704875
0.7 0.0477569302629697
0.8 0.0620918698200622
0.9 0.614042044807232
};
\addplot [thick, color1]
table {%
0.01 0.0165480901056542
0.1 0.0204934504911391
0.2 0.0206972216070032
0.3 0.0207879135837482
0.4 0.0208163095900146
0.5 0.0208476993949092
0.6 0.0208698401509094
0.7 0.0208743173128295
0.8 0.0208862758516474
0.9 0.020887968309707
};
\addplot [thick, color1, dashed]
table {%
0.01 0.0472313954386789
0.1 0.0275450666689814
0.2 0.0286040230035989
0.3 0.0357936352271847
0.4 0.0421961903357434
0.5 0.047980375883662
0.6 0.056379424622836
0.7 0.0560475251567388
0.8 0.044521677610201
0.9 0.0381787938864558
};
\addplot [thick, color1, densely dotted]
table {%
0.01 1.12881035380386
0.1 0.136674563237377
0.2 0.0814144502405203
0.3 0.0634084537839828
0.4 0.0529400006516673
0.5 0.0481356712450485
0.6 0.0432481510388553
0.7 0.0378517719547796
0.8 0.0366716194361524
0.9 0.0366986731823497
};
\end{axis}

\end{tikzpicture}

%% file: img/burst/norm_thr.tex
\begin{tikzpicture}

\definecolor{color0}{rgb}{0.12156862745098,0.466666666666667,0.705882352941177}
\definecolor{color1}{rgb}{0.83921568627451,0.152941176470588,0.156862745098039}

\begin{axis}[
width=\fwidth,
height=\fheight,
legend cell align={left},
legend style={fill opacity=0.8, draw opacity=1, text opacity=1, at={(0.03,0.03)}, anchor=south west, draw=white!80!black},
tick align=outside,
tick pos=left,
x grid style={white!69.0196078431373!black},
xlabel={$\eta$},
xmajorgrids,
xmin=-0.0395, xmax=1.0495,
xtick style={color=black},
xtick={-0.2,0,0.2,0.4,0.6,0.8,1,1.2},
xticklabels={−0.2,0.0,0.2,0.4,0.6,0.8,1.0,1.2},
y grid style={white!69.0196078431373!black},
ylabel={Norm. Aggr. Throughput},
ymajorgrids,
ymin=0.89, ymax=1.01250024380453,
]
\path [draw=color0, thick]
(axis cs:0.01,0.9999985168574)
--(axis cs:0.01,0.9999985168574);

\path [draw=color0, thick]
(axis cs:0.1,0.9999985168574)
--(axis cs:0.1,0.9999985168574);

\path [draw=color0, thick]
(axis cs:0.2,0.999999927332332)
--(axis cs:0.2,0.999999927332332);

\path [draw=color0, thick]
(axis cs:0.3,0.999999457174021)
--(axis cs:0.3,0.999999457174021);

\path [draw=color0, thick]
(axis cs:0.4,0.999999927332332)
--(axis cs:0.4,0.999999927332332);

\path [draw=color0, thick]
(axis cs:0.5,0.999996815972188)
--(axis cs:0.5,1.00000015162351);

\path [draw=color0, thick]
(axis cs:0.6,0.999996041963914)
--(axis cs:0.6,1.00000022959952);

\path [draw=color0, thick]
(axis cs:0.7,0.999998976697365)
--(axis cs:0.7,0.999999963102856);

\path [draw=color0, thick]
(axis cs:0.8,0.999998005299242)
--(axis cs:0.8,1.00000005781481);

\path [draw=color0, thick]
(axis cs:0.9,0.925817243964983)
--(axis cs:0.9,0.936186502624277);


\path [draw=color1, thick]
(axis cs:0.01,0.9999985168574)
--(axis cs:0.01,0.9999985168574);

\path [draw=color1, thick]
(axis cs:0.1,0.9999985168574)
--(axis cs:0.1,0.9999985168574);

\path [draw=color1, thick]
(axis cs:0.2,0.999999927332332)
--(axis cs:0.2,0.999999927332332);

\path [draw=color1, thick]
(axis cs:0.3,0.999999457174021)
--(axis cs:0.3,0.999999457174021);

\path [draw=color1, thick]
(axis cs:0.4,0.999999927332332)
--(axis cs:0.4,0.999999927332332);

\path [draw=color1, thick]
(axis cs:0.5,0.999999645237345)
--(axis cs:0.5,0.999999645237345);

\path [draw=color1, thick]
(axis cs:0.6,0.999999927332332)
--(axis cs:0.6,0.999999927332332);

\path [draw=color1, thick]
(axis cs:0.7,0.999999725835913)
--(axis cs:0.7,0.999999725835913);

\path [draw=color1, thick]
(axis cs:0.8,0.999999927332332)
--(axis cs:0.8,0.999999927332332);

\path [draw=color1, thick]
(axis cs:0.9,0.999999770612895)
--(axis cs:0.9,0.999999770612895);


\path [draw=color1, thick]
(axis cs:0.01,0.9999985168574)
--(axis cs:0.01,0.9999985168574);

\path [draw=color1, thick]
(axis cs:0.1,0.9999985168574)
--(axis cs:0.1,0.9999985168574);

\path [draw=color1, thick]
(axis cs:0.2,0.999999927332332)
--(axis cs:0.2,0.999999927332332);

\path [draw=color1, thick]
(axis cs:0.3,0.999999457174021)
--(axis cs:0.3,0.999999457174021);

\path [draw=color1, thick]
(axis cs:0.4,0.999999927332332)
--(axis cs:0.4,0.999999927332332);

\path [draw=color1, thick]
(axis cs:0.5,0.999996292981139)
--(axis cs:0.5,1.00000013101257);

\path [draw=color1, thick]
(axis cs:0.6,0.999998179342388)
--(axis cs:0.6,1.0000004809552);

\path [draw=color1, thick]
(axis cs:0.7,0.999998528601553)
--(axis cs:0.7,0.999999899327064);

\path [draw=color1, thick]
(axis cs:0.8,0.999998879752267)
--(axis cs:0.8,1.00000007913709);

\path [draw=color1, thick]
(axis cs:0.9,0.975607081374598)
--(axis cs:0.9,1.00772460264057);


\path [draw=color1, thick]
(axis cs:0.01,0.9999985168574)
--(axis cs:0.01,0.9999985168574);

\path [draw=color1, thick]
(axis cs:0.1,0.9999985168574)
--(axis cs:0.1,0.9999985168574);

\path [draw=color1, thick]
(axis cs:0.2,0.999999927332332)
--(axis cs:0.2,0.999999927332332);

\path [draw=color1, thick]
(axis cs:0.3,0.999999457174021)
--(axis cs:0.3,0.999999457174021);

\path [draw=color1, thick]
(axis cs:0.4,0.999999927332332)
--(axis cs:0.4,0.999999927332332);

\path [draw=color1, thick]
(axis cs:0.5,0.999999645237345)
--(axis cs:0.5,0.999999645237345);

\path [draw=color1, thick]
(axis cs:0.6,0.999999927332332)
--(axis cs:0.6,0.999999927332332);

\path [draw=color1, thick]
(axis cs:0.7,0.999999725835913)
--(axis cs:0.7,0.999999725835913);

\path [draw=color1, thick]
(axis cs:0.8,0.999999927332332)
--(axis cs:0.8,0.999999927332332);

\path [draw=color1, thick]
(axis cs:0.9,0.999999770612895)
--(axis cs:0.9,0.999999770612895);


\addplot [thick, color0]
table {%
0.01 0.9999985168574
0.1 0.9999985168574
0.2 0.999999927332332
0.3 0.999999457174021
0.4 0.999999927332332
0.5 0.999998483797847
0.6 0.999998135781717
0.7 0.999999469900111
0.8 0.999999031557024
0.9 0.93100187329463
};
\addplot [thick, color1]
table {%
0.01 0.9999985168574
0.1 0.9999985168574
0.2 0.999999927332332
0.3 0.999999457174021
0.4 0.999999927332332
0.5 0.999999645237345
0.6 0.999999927332332
0.7 0.999999725835913
0.8 0.999999927332332
0.9 0.999999770612895
};
\addplot [thick, color1, dashed]
table {%
0.01 0.9999985168574
0.1 0.9999985168574
0.2 0.999999927332332
0.3 0.999999457174021
0.4 0.999999927332332
0.5 0.999998211996854
0.6 0.999999330148794
0.7 0.999999213964309
0.8 0.999999479444678
0.9 0.991665842007582
};
\addplot [thick, color1, densely dotted]
table {%
0.01 0.9999985168574
0.1 0.9999985168574
0.2 0.999999927332332
0.3 0.999999457174021
0.4 0.999999927332332
0.5 0.999999645237345
0.6 0.999999927332332
0.7 0.999999725835913
0.8 0.999999927332332
0.9 0.999999770612895
};
\end{axis}

\end{tikzpicture}

%% file: img/50Mbps/aggr_thr.tex
\begin{tikzpicture}

\definecolor{color0}{rgb}{0.12156862745098,0.466666666666667,0.705882352941177}
\definecolor{color1}{rgb}{0.83921568627451,0.152941176470588,0.156862745098039}

\begin{axis}[
width=\fwidth,
height=\fheight,
legend cell align={left},
legend style={fill opacity=0.8, draw opacity=1, text opacity=1, at={(0.03,0.97)}, anchor=north west, draw=white!80!black},
tick align=outside,
tick pos=left,
x grid style={white!69.0196078431373!black},
xlabel={\# STAs},
xmajorgrids,
minor x tick num=1,
xminorgrids,
xmin=0, xmax=10.45,
xtick style={color=black},
y grid style={white!69.0196078431373!black},
ylabel={Aggr. Throughput [Mbps]},
ymajorgrids,
ymin=0, ymax=1045,
ytick style={color=black}
]
\path [draw=color0, thick]
(axis cs:1,50)
--(axis cs:1,50);

\path [draw=color0, thick]
(axis cs:2,100)
--(axis cs:2,100);

\path [draw=color0, thick]
(axis cs:3,150)
--(axis cs:3,150);

\path [draw=color0, thick]
(axis cs:4,200)
--(axis cs:4,200);

\path [draw=color0, thick]
(axis cs:5,250)
--(axis cs:5,250);

\path [draw=color0, thick]
(axis cs:6,300)
--(axis cs:6,300);

\path [draw=color0, thick]
(axis cs:7,350)
--(axis cs:7,350);

\path [draw=color0, thick]
(axis cs:8,400)
--(axis cs:8,400);

\path [draw=color0, thick]
(axis cs:9,450)
--(axis cs:9,450);

\path [draw=color0, thick]
(axis cs:10,500)
--(axis cs:10,500);

\path [draw=color1, thick]
(axis cs:1,50)
--(axis cs:1,50);

\path [draw=color1, thick]
(axis cs:2,100)
--(axis cs:2,100);

\path [draw=color1, thick]
(axis cs:3,150)
--(axis cs:3,150);

\path [draw=color1, thick]
(axis cs:4,200)
--(axis cs:4,200);

\path [draw=color1, thick]
(axis cs:5,250)
--(axis cs:5,250);

\path [draw=color1, thick]
(axis cs:6,300)
--(axis cs:6,300);

\path [draw=color1, thick]
(axis cs:7,350)
--(axis cs:7,350);

\path [draw=color1, thick]
(axis cs:8,400)
--(axis cs:8,400);

\path [draw=color1, thick]
(axis cs:9,450)
--(axis cs:9,450);

\path [draw=color1, thick]
(axis cs:10,500)
--(axis cs:10,500);

\path [draw=color1, thick]
(axis cs:1,50)
--(axis cs:1,50);

\path [draw=color1, thick]
(axis cs:2,100)
--(axis cs:2,100);

\path [draw=color1, thick]
(axis cs:3,150)
--(axis cs:3,150);

\path [draw=color1, thick]
(axis cs:4,200)
--(axis cs:4,200);

\path [draw=color1, thick]
(axis cs:5,250)
--(axis cs:5,250);

\path [draw=color1, thick]
(axis cs:6,300)
--(axis cs:6,300);

\path [draw=color1, thick]
(axis cs:7,350)
--(axis cs:7,350);

\path [draw=color1, thick]
(axis cs:8,400)
--(axis cs:8,400);

\path [draw=color1, thick]
(axis cs:9,449.998838163986)
--(axis cs:9,450.000367976365);

\path [draw=color1, thick]
(axis cs:10,499.997767368709)
--(axis cs:10,500.000247982169);

\path [draw=color1, thick]
(axis cs:1,50)
--(axis cs:1,50);

\path [draw=color1, thick]
(axis cs:2,100)
--(axis cs:2,100);

\path [draw=color1, thick]
(axis cs:3,150)
--(axis cs:3,150);

\path [draw=color1, thick]
(axis cs:4,200)
--(axis cs:4,200);

\path [draw=color1, thick]
(axis cs:5,250)
--(axis cs:5,250);

\path [draw=color1, thick]
(axis cs:6,300)
--(axis cs:6,300);

\path [draw=color1, thick]
(axis cs:7,350)
--(axis cs:7,350);

\path [draw=color1, thick]
(axis cs:8,400)
--(axis cs:8,400);

\path [draw=color1, thick]
(axis cs:9,450)
--(axis cs:9,450);

\path [draw=color1, thick]
(axis cs:10,500)
--(axis cs:10,500);

\addplot [thick, color0]
table {%
1 50
2 100
3 150
4 200
5 250
6 300
7 350
8 400
9 450
10 500
};
\addplot [thick, color1]
table {%
1 50
2 100
3 150
4 200
5 250
6 300
7 350
8 400
9 450
10 500
};
\addplot [thick, color1, dashed]
table {%
1 50
2 100
3 150
4 200
5 250
6 300
7 350
8 400
9 449.999603070176
10 499.999007675439
};
\addplot [thick, color1, densely dotted]
table {%
1 50
2 100
3 150
4 200
5 250
6 300
7 350
8 400
9 450
10 500
};
\end{axis}

\end{tikzpicture}

%% file: img/100Mbps/aggr_thr.tex
\begin{tikzpicture}

\definecolor{color0}{rgb}{0.12156862745098,0.466666666666667,0.705882352941177}
\definecolor{color1}{rgb}{0.83921568627451,0.152941176470588,0.156862745098039}

\begin{axis}[
width=\fwidth,
height=\fheight,
legend cell align={left},
legend style={fill opacity=0.8, draw opacity=1, text opacity=1, at={(0.03,0.97)}, anchor=north west, draw=white!80!black},
tick align=outside,
tick pos=left,
x grid style={white!69.0196078431373!black},
xlabel={\# STAs},
xmajorgrids,
minor x tick num=1,
xminorgrids,
xmin=0, xmax=10.45,
xtick style={color=black},
y grid style={white!69.0196078431373!black},
ylabel={Aggr. Throughput [Mbps]},
ymajorgrids,
ymin=0, ymax=1045,
ytick style={color=black}
]
\path [draw=color0, thick]
(axis cs:1,100)
--(axis cs:1,100);

\path [draw=color0, thick]
(axis cs:2,200)
--(axis cs:2,200);

\path [draw=color0, thick]
(axis cs:3,300)
--(axis cs:3,300);

\path [draw=color0, thick]
(axis cs:4,400)
--(axis cs:4,400);

\path [draw=color0, thick]
(axis cs:5,500)
--(axis cs:5,500);

\path [draw=color0, thick]
(axis cs:6,599.998765427452)
--(axis cs:6,600.000043783074);

\path [draw=color0, thick]
(axis cs:7,699.998765427452)
--(axis cs:7,700.000043783074);

\path [draw=color0, thick]
(axis cs:8,799.995058716901)
--(axis cs:8,799.99898733573);

\path [draw=color0, thick]
(axis cs:9,885.15338150784)
--(axis cs:9,893.913968053563);

\path [draw=color0, thick]
(axis cs:10,858.414822929971)
--(axis cs:10,862.479142859503);

\path [draw=color1, thick]
(axis cs:1,100)
--(axis cs:1,100);

\path [draw=color1, thick]
(axis cs:2,200)
--(axis cs:2,200);

\path [draw=color1, thick]
(axis cs:3,300)
--(axis cs:3,300);

\path [draw=color1, thick]
(axis cs:4,400)
--(axis cs:4,400);

\path [draw=color1, thick]
(axis cs:5,500)
--(axis cs:5,500);

\path [draw=color1, thick]
(axis cs:6,600)
--(axis cs:6,600);

\path [draw=color1, thick]
(axis cs:7,700)
--(axis cs:7,700);

\path [draw=color1, thick]
(axis cs:8,800)
--(axis cs:8,800);

\path [draw=color1, thick]
(axis cs:9,900)
--(axis cs:9,900);

\path [draw=color1, thick]
(axis cs:10,1000)
--(axis cs:10,1000);

\path [draw=color1, thick]
(axis cs:1,100)
--(axis cs:1,100);

\path [draw=color1, thick]
(axis cs:2,200)
--(axis cs:2,200);

\path [draw=color1, thick]
(axis cs:3,300)
--(axis cs:3,300);

\path [draw=color1, thick]
(axis cs:4,399.999419081993)
--(axis cs:4,400.000183988182);

\path [draw=color1, thick]
(axis cs:5,499.998296643699)
--(axis cs:5,500.000115637003);

\path [draw=color1, thick]
(axis cs:6,599.994578254356)
--(axis cs:6,599.999467798275);

\path [draw=color1, thick]
(axis cs:7,699.993750845537)
--(axis cs:7,699.998075908849);

\path [draw=color1, thick]
(axis cs:8,799.996246973195)
--(axis cs:8,799.999783728559);

\path [draw=color1, thick]
(axis cs:9,890.179943855172)
--(axis cs:9,903.019240355355);

\path [draw=color1, thick]
(axis cs:10,847.76299954261)
--(axis cs:10,937.907752650372);

\path [draw=color1, thick]
(axis cs:1,100)
--(axis cs:1,100);

\path [draw=color1, thick]
(axis cs:2,200)
--(axis cs:2,200);

\path [draw=color1, thick]
(axis cs:3,300)
--(axis cs:3,300);

\path [draw=color1, thick]
(axis cs:4,400)
--(axis cs:4,400);

\path [draw=color1, thick]
(axis cs:5,500)
--(axis cs:5,500);

\path [draw=color1, thick]
(axis cs:6,600)
--(axis cs:6,600);

\path [draw=color1, thick]
(axis cs:7,700)
--(axis cs:7,700);

\path [draw=color1, thick]
(axis cs:8,800)
--(axis cs:8,800);

\path [draw=color1, thick]
(axis cs:9,900)
--(axis cs:9,900);

\path [draw=color1, thick]
(axis cs:10,1000)
--(axis cs:10,1000);

\addplot [thick, color0]
table {%
1 100
2 200
3 300
4 400
5 500
6 599.999404605263
7 699.999404605263
8 799.997023026316
9 889.533674780702
10 860.446982894737
};
\addplot [thick, color1]
table {%
1 100
2 200
3 300
4 400
5 500
6 600
7 700
8 800
9 900
10 1000
};
\addplot [thick, color1, dashed]
table {%
1 100
2 200
3 300
4 399.999801535088
5 499.999206140351
6 599.997023026316
7 699.995913377193
8 799.998015350877
9 896.599592105263
10 892.835376096491
};
\addplot [thick, color1, densely dotted]
table {%
1 100
2 200
3 300
4 400
5 500
6 600
7 700
8 800
9 900
10 1000
};
\end{axis}

\end{tikzpicture}

%% file: img/200Mbps/aggr_thr.tex
\begin{tikzpicture}

\definecolor{color0}{rgb}{0.12156862745098,0.466666666666667,0.705882352941177}
\definecolor{color1}{rgb}{0.83921568627451,0.152941176470588,0.156862745098039}

\begin{axis}[
width=\fwidth,
height=\fheight,
legend cell align={left},
legend style={fill opacity=0.8, draw opacity=1, text opacity=1, at={(0.03,0.97)}, anchor=north west, draw=white!80!black},
tick align=outside,
tick pos=left,
x grid style={white!69.0196078431373!black},
xlabel={\# STAs},
xmajorgrids,
minor x tick num=1,
xminorgrids,
xmin=0, xmax=10.45,
xtick style={color=black},
y grid style={white!69.0196078431373!black},
ylabel={Aggr. Throughput [Mbps]},
ymajorgrids,
ymin=0, ymax=1040,
ytick style={color=black}
]
\path [draw=color0, thick]
(axis cs:1,200)
--(axis cs:1,200);

\path [draw=color0, thick]
(axis cs:2,400)
--(axis cs:2,400);

\path [draw=color0, thick]
(axis cs:3,599.998009294136)
--(axis cs:3,600.000402986566);

\path [draw=color0, thick]
(axis cs:4,799.999419081993)
--(axis cs:4,800.000183988182);

\path [draw=color0, thick]
(axis cs:5,906.101857896423)
--(axis cs:5,918.570110086034);

\path [draw=color0, thick]
(axis cs:6,883.112290647338)
--(axis cs:6,887.144534791259);

\path [draw=color0, thick]
(axis cs:7,867.260695344798)
--(axis cs:7,871.160171321869);

\path [draw=color0, thick]
(axis cs:8,853.410138198045)
--(axis cs:8,858.950271451078);

\path [draw=color0, thick]
(axis cs:9,841.348266802378)
--(axis cs:9,846.381245039727);

\path [draw=color0, thick]
(axis cs:10,830.721126275302)
--(axis cs:10,836.746078110663);

\path [draw=color1, thick]
(axis cs:1,200)
--(axis cs:1,200);

\path [draw=color1, thick]
(axis cs:2,400)
--(axis cs:2,400);

\path [draw=color1, thick]
(axis cs:3,600)
--(axis cs:3,600);

\path [draw=color1, thick]
(axis cs:4,800)
--(axis cs:4,800);

\path [draw=color1, thick]
(axis cs:5,1000)
--(axis cs:5,1000);

\path [draw=color1, thick]
(axis cs:6,1000)
--(axis cs:6,1000);

\path [draw=color1, thick]
(axis cs:7,1000)
--(axis cs:7,1000);

\path [draw=color1, thick]
(axis cs:8,1000)
--(axis cs:8,1000);

\path [draw=color1, thick]
(axis cs:9,1000)
--(axis cs:9,1000);

\path [draw=color1, thick]
(axis cs:10,1000)
--(axis cs:10,1000);

\path [draw=color1, thick]
(axis cs:1,200)
--(axis cs:1,200);

\path [draw=color1, thick]
(axis cs:2,400)
--(axis cs:2,400);

\path [draw=color1, thick]
(axis cs:3,599.998296643699)
--(axis cs:3,600.000115637003);

\path [draw=color1, thick]
(axis cs:4,799.99907160698)
--(axis cs:4,800.00013453337);

\path [draw=color1, thick]
(axis cs:5,980.486104868837)
--(axis cs:5,1006.180164868);

\path [draw=color1, thick]
(axis cs:6,999.998561292377)
--(axis cs:6,1000.00024791815);

\path [draw=color1, thick]
(axis cs:7,999.94226188529)
--(axis cs:7,1000.01828679892);

\path [draw=color1, thick]
(axis cs:8,999.99907160698)
--(axis cs:8,1000.00013453337);

\path [draw=color1, thick]
(axis cs:9,999.998257245979)
--(axis cs:9,1000.00055196455);

\path [draw=color1, thick]
(axis cs:10,958.112036326065)
--(axis cs:10,1000.98076542832);

\path [draw=color1, thick]
(axis cs:1,200)
--(axis cs:1,200);

\path [draw=color1, thick]
(axis cs:2,400)
--(axis cs:2,400);

\path [draw=color1, thick]
(axis cs:3,600)
--(axis cs:3,600);

\path [draw=color1, thick]
(axis cs:4,800)
--(axis cs:4,800);

\path [draw=color1, thick]
(axis cs:5,1000)
--(axis cs:5,1000);

\path [draw=color1, thick]
(axis cs:6,1000)
--(axis cs:6,1000);

\path [draw=color1, thick]
(axis cs:7,1000)
--(axis cs:7,1000);

\path [draw=color1, thick]
(axis cs:8,1000)
--(axis cs:8,1000);

\path [draw=color1, thick]
(axis cs:9,1000)
--(axis cs:9,1000);

\path [draw=color1, thick]
(axis cs:10,1000)
--(axis cs:10,1000);

\addplot [thick, color0]
table {%
1 200
2 400
3 599.999206140351
4 799.999801535088
5 912.335983991228
6 885.128412719298
7 869.210433333333
8 856.180204824562
9 843.864755921053
10 833.733602192983
};
\addplot [thick, color1]
table {%
1 200
2 400
3 600
4 800
5 1000
6 1000
7 1000
8 1000
9 1000
10 1000
};
\addplot [thick, color1, dashed]
table {%
1 200
2 400
3 599.999206140351
4 799.999603070175
5 993.333134868421
6 999.999404605263
7 999.980274342105
8 999.999603070175
9 999.999404605263
10 979.546400877193
};
\addplot [thick, color1, densely dotted]
table {%
1 200
2 400
3 600
4 800
5 1000
6 1000
7 1000
8 1000
9 1000
10 1000
};
\end{axis}

\end{tikzpicture}

%% file: img/50Mbps/avg_delay.tex
\begin{tikzpicture}

\definecolor{color0}{rgb}{0.12156862745098,0.466666666666667,0.705882352941177}
\definecolor{color1}{rgb}{0.83921568627451,0.152941176470588,0.156862745098039}

\begin{axis}[
width=\fwidth,
height=\fheight,
legend cell align={left},
legend style={fill opacity=0.8, draw opacity=1, text opacity=1, draw=white!80!black},
tick align=outside,
tick pos=left,
x grid style={white!69.0196078431373!black},
xlabel={\# STAs},
xmajorgrids,
minor x tick num=1,
xminorgrids,
xmin=0, xmax=10.45,
xtick style={color=black},
y grid style={white!69.0196078431373!black},
ylabel={Avg. Delay [ms]},
ymajorgrids,
ymin=0, ymax=60,
ytick style={color=black}
]
\path [draw=color0, thick]
(axis cs:1,2.79356153998679)
--(axis cs:1,4.18592205237654);

\path [draw=color0, thick]
(axis cs:2,3.17920201473434)
--(axis cs:2,4.00994213204662);

\path [draw=color0, thick]
(axis cs:3,3.0868702947312)
--(axis cs:3,3.99321103010126);

\path [draw=color0, thick]
(axis cs:4,2.92434234841176)
--(axis cs:4,3.60803446506423);

\path [draw=color0, thick]
(axis cs:5,3.01972123850486)
--(axis cs:5,3.69323288845714);

\path [draw=color0, thick]
(axis cs:6,3.06556715360125)
--(axis cs:6,4.15723105964246);

\path [draw=color0, thick]
(axis cs:7,3.48046134610381)
--(axis cs:7,4.72218436970603);

\path [draw=color0, thick]
(axis cs:8,4.15219792738354)
--(axis cs:8,5.14475119970507);

\path [draw=color0, thick]
(axis cs:9,3.92328803948567)
--(axis cs:9,4.9713948565914);

\path [draw=color0, thick]
(axis cs:10,5.09155548780213)
--(axis cs:10,6.29904086383299);

\path [draw=color1, thick]
(axis cs:1,2.3391039805279)
--(axis cs:1,2.33915154008297);

\path [draw=color1, thick]
(axis cs:2,2.33962728006406)
--(axis cs:2,2.33987210110211);

\path [draw=color1, thick]
(axis cs:3,2.33994841489419)
--(axis cs:3,2.34040327495611);

\path [draw=color1, thick]
(axis cs:4,2.34028717536326)
--(axis cs:4,2.34081792864341);

\path [draw=color1, thick]
(axis cs:5,2.34028739549851)
--(axis cs:5,2.34077859092429);

\path [draw=color1, thick]
(axis cs:6,2.34044556247848)
--(axis cs:6,2.34110895906152);

\path [draw=color1, thick]
(axis cs:7,2.34044313444165)
--(axis cs:7,2.3410838257789);

\path [draw=color1, thick]
(axis cs:8,2.34062041123517)
--(axis cs:8,2.34125550397933);

\path [draw=color1, thick]
(axis cs:9,2.3405182875695)
--(axis cs:9,2.34122997355741);

\path [draw=color1, thick]
(axis cs:10,2.34068497837621)
--(axis cs:10,2.34101947358324);

\path [draw=color1, thick]
(axis cs:1,2.43592858485647)
--(axis cs:1,3.33865959719803);

\path [draw=color1, thick]
(axis cs:2,3.34456699993637)
--(axis cs:2,4.51532196216018);

\path [draw=color1, thick]
(axis cs:3,3.64312727254574)
--(axis cs:3,5.00745320520336);

\path [draw=color1, thick]
(axis cs:4,4.15061306114073)
--(axis cs:4,5.5106909988526);

\path [draw=color1, thick]
(axis cs:5,5.1306850979472)
--(axis cs:5,6.75876921670374);

\path [draw=color1, thick]
(axis cs:6,5.76311240885481)
--(axis cs:6,7.79868364774518);

\path [draw=color1, thick]
(axis cs:7,7.73994185126539)
--(axis cs:7,9.36190427540553);

\path [draw=color1, thick]
(axis cs:8,7.91263312624052)
--(axis cs:8,10.4581684707338);

\path [draw=color1, thick]
(axis cs:9,11.9612557432604)
--(axis cs:9,14.64526849916);

\path [draw=color1, thick]
(axis cs:10,13.7315070811988)
--(axis cs:10,16.9298273032111);

\path [draw=color1, thick]
(axis cs:1,36.028377697536)
--(axis cs:1,58.4758847351412);

\path [draw=color1, thick]
(axis cs:2,34.7593499313823)
--(axis cs:2,48.8353744234609);

\path [draw=color1, thick]
(axis cs:3,44.5950199402668)
--(axis cs:3,55.9526032417651);

\path [draw=color1, thick]
(axis cs:4,47.5565428335531)
--(axis cs:4,59.2355715751049);

\path [draw=color1, thick]
(axis cs:5,43.6101301070272)
--(axis cs:5,52.4640635633647);

\path [draw=color1, thick]
(axis cs:6,44.3331423393392)
--(axis cs:6,51.6263229608029);

\path [draw=color1, thick]
(axis cs:7,46.453444199771)
--(axis cs:7,55.0393848908655);

\path [draw=color1, thick]
(axis cs:8,47.5448455812219)
--(axis cs:8,53.7197596998596);

\path [draw=color1, thick]
(axis cs:9,49.4553515286652)
--(axis cs:9,55.0921837612713);

\path [draw=color1, thick]
(axis cs:10,47.5361655174568)
--(axis cs:10,54.0073692002505);

\addplot [thick, color0]
table {%
1 3.48974179618167
2 3.59457207339048
3 3.54004066241623
4 3.26618840673799
5 3.356477063481
6 3.61139910662186
7 4.10132285790492
8 4.64847456354431
9 4.44734144803853
10 5.69529817581756
};
\addplot [thick, color1]
table {%
1 2.33912776030544
2 2.33974969058308
3 2.34017584492515
4 2.34055255200333
5 2.3405329932114
6 2.34077726077
7 2.34076348011028
8 2.34093795760725
9 2.34087413056346
10 2.34085222597973
};
\addplot [thick, color1, dashed]
table {%
1 2.88729409102725
2 3.92994448104827
3 4.32529023887455
4 4.83065202999667
5 5.94472715732547
6 6.78089802829999
7 8.55092306333546
8 9.18540079848716
9 13.3032621212102
10 15.3306671922049
};
\addplot [thick, color1, densely dotted]
table {%
1 47.2521312163386
2 41.7973621774216
3 50.273811591016
4 53.396057204329
5 48.037096835196
6 47.979732650071
7 50.7464145453182
8 50.6323026405408
9 52.2737676449683
10 50.7717673588537
};
\end{axis}

\end{tikzpicture}

%% file: img/100Mbps/avg_delay.tex
\begin{tikzpicture}

\definecolor{color0}{rgb}{0.12156862745098,0.466666666666667,0.705882352941177}
\definecolor{color1}{rgb}{0.83921568627451,0.152941176470588,0.156862745098039}

\begin{axis}[
width=\fwidth,
height=\fheight,
legend cell align={left},
legend style={fill opacity=0.8, draw opacity=1, text opacity=1, at={(0.03,0.97)}, anchor=north west, draw=white!80!black},
tick align=outside,
tick pos=left,
x grid style={white!69.0196078431373!black},
xlabel={\# STAs},
xmajorgrids,
minor x tick num=1,
xminorgrids,
xmin=0, xmax=10.45,
xtick style={color=black},
y grid style={white!69.0196078431373!black},
ylabel={Avg. Delay [ms]},
ymajorgrids,
ymin=0, ymax=60,
ytick style={color=black}
]
\path [draw=color0, thick]
(axis cs:1,5.48253120769875)
--(axis cs:1,7.0384919644539);

\path [draw=color0, thick]
(axis cs:2,5.91406327889344)
--(axis cs:2,6.82386170803819);

\path [draw=color0, thick]
(axis cs:3,6.24869000978456)
--(axis cs:3,7.91916383971478);

\path [draw=color0, thick]
(axis cs:4,6.88349170509764)
--(axis cs:4,8.55755728071343);

\path [draw=color0, thick]
(axis cs:5,6.95582378833184)
--(axis cs:5,8.58744998659848);

\path [draw=color0, thick]
(axis cs:6,8.48174172441071)
--(axis cs:6,11.0870466046998);

\path [draw=color0, thick]
(axis cs:7,9.95991147576339)
--(axis cs:7,12.8602373838615);

\path [draw=color0, thick]
(axis cs:8,12.376731524307)
--(axis cs:8,16.1881497045908);

\path [draw=color0, thick]
(axis cs:9,31.4701314487796)
--(axis cs:9,54.0337428274672);

\path [draw=color0, thick]
(axis cs:10,134.719817039801)
--(axis cs:10,142.44561001631);

\path [draw=color1, thick]
(axis cs:1,4.65556166970266)
--(axis cs:1,4.65558546848533);

\path [draw=color1, thick]
(axis cs:2,4.65582205080709)
--(axis cs:2,4.65594499616605);

\path [draw=color1, thick]
(axis cs:3,4.6559828237825)
--(axis cs:3,4.65621114921318);

\path [draw=color1, thick]
(axis cs:4,4.65615102532074)
--(axis cs:4,4.65641797875375);

\path [draw=color1, thick]
(axis cs:5,4.65614867145751)
--(axis cs:5,4.65639580823064);

\path [draw=color1, thick]
(axis cs:6,4.65622561039635)
--(axis cs:6,4.65655960593137);

\path [draw=color1, thick]
(axis cs:7,4.6562232825988)
--(axis cs:7,4.65654583348579);

\path [draw=color1, thick]
(axis cs:8,4.65630972143047)
--(axis cs:8,4.65662945609822);

\path [draw=color1, thick]
(axis cs:9,4.65625506953328)
--(axis cs:9,4.65661330516807);

\path [draw=color1, thick]
(axis cs:10,4.65633910283229)
--(axis cs:10,4.65650838533203);

\path [draw=color1, thick]
(axis cs:1,4.95590595027379)
--(axis cs:1,6.14305784407168);

\path [draw=color1, thick]
(axis cs:2,6.68308686492436)
--(axis cs:2,8.39841998527741);

\path [draw=color1, thick]
(axis cs:3,8.21104381513749)
--(axis cs:3,11.0533474833281);

\path [draw=color1, thick]
(axis cs:4,11.8971482273952)
--(axis cs:4,14.8753713289326);

\path [draw=color1, thick]
(axis cs:5,14.2313745930353)
--(axis cs:5,17.4688181606576);

\path [draw=color1, thick]
(axis cs:6,20.7341960455436)
--(axis cs:6,24.857518263985);

\path [draw=color1, thick]
(axis cs:7,28.7989896314526)
--(axis cs:7,34.4870464891623);

\path [draw=color1, thick]
(axis cs:8,38.7126623682908)
--(axis cs:8,45.4698233774666);

\path [draw=color1, thick]
(axis cs:9,43.425955623546)
--(axis cs:9,49.0190288936767);

\path [draw=color1, thick]
(axis cs:10,48.5454670576197)
--(axis cs:10,56.7297463771111);

\path [draw=color1, thick]
(axis cs:1,38.2949916270692)
--(axis cs:1,60.584798427271);

\path [draw=color1, thick]
(axis cs:2,36.2020298135779)
--(axis cs:2,48.890421778351);

\path [draw=color1, thick]
(axis cs:3,45.1922582271304)
--(axis cs:3,54.9513503652521);

\path [draw=color1, thick]
(axis cs:4,46.3785890004964)
--(axis cs:4,56.317123635481);

\path [draw=color1, thick]
(axis cs:5,43.4801504100988)
--(axis cs:5,51.9685889717286);

\path [draw=color1, thick]
(axis cs:6,45.8523289900413)
--(axis cs:6,52.5620194940259);

\path [draw=color1, thick]
(axis cs:7,46.6742575485918)
--(axis cs:7,53.20875383474);

\path [draw=color1, thick]
(axis cs:8,49.6346151680057)
--(axis cs:8,57.0191902922864);

\path [draw=color1, thick]
(axis cs:9,47.7893956320837)
--(axis cs:9,53.4624806807893);

\path [draw=color1, thick]
(axis cs:10,48.0404447564535)
--(axis cs:10,53.8225619693865);

\addplot [thick, color0]
table {%
1 6.26051158607633
2 6.36896249346582
3 7.08392692474967
4 7.72052449290553
5 7.77163688746516
6 9.78439416455526
7 11.4100744298124
8 14.2824406144489
9 42.7519371381234
10 138.582713528056
};
\addplot [thick, color1]
table {%
1 4.655573569094
2 4.65588352348657
3 4.65609698649784
4 4.65628450203724
5 4.65627223984408
6 4.65639260816386
7 4.65638455804229
8 4.65646958876434
9 4.65643418735068
10 4.65642374408216
};
\addplot [thick, color1, dashed]
table {%
1 5.54948189717273
2 7.54075342510089
3 9.6321956492328
4 13.3862597781639
5 15.8500963768465
6 22.7958571547643
7 31.6430180603074
8 42.0912428728787
9 46.2224922586114
10 52.6376067173654
};
\addplot [thick, color1, densely dotted]
table {%
1 49.4398950271701
2 42.5462257959645
3 50.0718042961913
4 51.3478563179887
5 47.7243696909137
6 49.2071742420336
7 49.9415056916659
8 53.3269027301461
9 50.6259381564365
10 50.93150336292
};
\end{axis}

\end{tikzpicture}

%% file: img/200Mbps/avg_delay.tex
\begin{tikzpicture}

\definecolor{color0}{rgb}{0.12156862745098,0.466666666666667,0.705882352941177}
\definecolor{color1}{rgb}{0.83921568627451,0.152941176470588,0.156862745098039}

\begin{axis}[
width=\fwidth,
height=\fheight,
legend cell align={left},
legend style={fill opacity=0.8, draw opacity=1, text opacity=1, at={(0.03,0.97)}, anchor=north west, draw=white!80!black},
tick align=outside,
tick pos=left,
x grid style={white!69.0196078431373!black},
xlabel={\# STAs},
xmajorgrids,
minor x tick num=1,
xminorgrids,
xmin=0, xmax=10.45,
xtick style={color=black},
y grid style={white!69.0196078431373!black},
ylabel={Avg. Delay [ms]},
ymajorgrids,
ymin=0, ymax=60,
ytick style={color=black}
]
\path [draw=color0, thick]
(axis cs:1,10.7018438620268)
--(axis cs:1,12.3815173334883);

\path [draw=color0, thick]
(axis cs:2,12.1080459300143)
--(axis cs:2,13.8653106594619);

\path [draw=color0, thick]
(axis cs:3,14.4469613368666)
--(axis cs:3,18.654380915224);

\path [draw=color0, thick]
(axis cs:4,18.963026308728)
--(axis cs:4,23.3505612302589);

\path [draw=color0, thick]
(axis cs:5,172.258841921037)
--(axis cs:5,203.339653373178);

\path [draw=color0, thick]
(axis cs:6,309.704211944599)
--(axis cs:6,327.600143357794);

\path [draw=color0, thick]
(axis cs:7,356.849314341866)
--(axis cs:7,368.699588630946);

\path [draw=color0, thick]
(axis cs:8,385.082538843516)
--(axis cs:8,393.684889880312);

\path [draw=color0, thick]
(axis cs:9,406.531416876631)
--(axis cs:9,412.213718293645);

\path [draw=color0, thick]
(axis cs:10,415.441392344394)
--(axis cs:10,420.877433059539);

\path [draw=color1, thick]
(axis cs:1,9.28992678404916)
--(axis cs:1,9.28993868819749);

\path [draw=color1, thick]
(axis cs:2,9.29005548001519)
--(axis cs:2,9.29011754997206);

\path [draw=color1, thick]
(axis cs:3,9.290136031775)
--(axis cs:3,9.2902509780109);

\path [draw=color1, thick]
(axis cs:4,9.29021885609569)
--(axis cs:4,9.2903537653741);

\path [draw=color1, thick]
(axis cs:5,9.29021519772269)
--(axis cs:5,9.29034020196836);

\path [draw=color1, thick]
(axis cs:6,9.29027138623382)
--(axis cs:6,9.29043292667658);

\path [draw=color1, thick]
(axis cs:7,9.29029494902251)
--(axis cs:7,9.29050100793259);

\path [draw=color1, thick]
(axis cs:8,9.29032227976501)
--(axis cs:8,9.29049688266116);

\path [draw=color1, thick]
(axis cs:9,9.29025296677162)
--(axis cs:9,9.29047209166547);

\path [draw=color1, thick]
(axis cs:10,9.29030631047066)
--(axis cs:10,9.29048562255292);

\path [draw=color1, thick]
(axis cs:1,10.2366500634064)
--(axis cs:1,11.6847624431442);

\path [draw=color1, thick]
(axis cs:2,14.303250289101)
--(axis cs:2,17.813774865317);

\path [draw=color1, thick]
(axis cs:3,21.3502464195755)
--(axis cs:3,28.6453102519626);

\path [draw=color1, thick]
(axis cs:4,36.7758157614097)
--(axis cs:4,43.5652352173196);

\path [draw=color1, thick]
(axis cs:5,42.7151017607912)
--(axis cs:5,49.9558746776974);

\path [draw=color1, thick]
(axis cs:6,47.6533969810205)
--(axis cs:6,55.0643040459541);

\path [draw=color1, thick]
(axis cs:7,47.0828004714537)
--(axis cs:7,55.2990053536925);

\path [draw=color1, thick]
(axis cs:8,44.4438908224327)
--(axis cs:8,53.3856454111965);

\path [draw=color1, thick]
(axis cs:9,45.4368199576735)
--(axis cs:9,52.2423616483141);

\path [draw=color1, thick]
(axis cs:10,44.7678227727386)
--(axis cs:10,56.2159609772844);

\path [draw=color1, thick]
(axis cs:1,39.2863054801843)
--(axis cs:1,57.3516245523063);

\path [draw=color1, thick]
(axis cs:2,37.8517887152487)
--(axis cs:2,49.1876859000383);

\path [draw=color1, thick]
(axis cs:3,44.1158133569246)
--(axis cs:3,54.5270262218098);

\path [draw=color1, thick]
(axis cs:4,47.3534925729287)
--(axis cs:4,54.9512917596635);

\path [draw=color1, thick]
(axis cs:5,42.8082175661759)
--(axis cs:5,49.6184792502603);

\path [draw=color1, thick]
(axis cs:6,48.2351361382872)
--(axis cs:6,55.3648943500533);

\path [draw=color1, thick]
(axis cs:7,46.6502527338549)
--(axis cs:7,54.5646805037846);

\path [draw=color1, thick]
(axis cs:8,45.2995684827952)
--(axis cs:8,54.0608512261861);

\path [draw=color1, thick]
(axis cs:9,45.9924701310925)
--(axis cs:9,52.5584626545942);

\path [draw=color1, thick]
(axis cs:10,45.4852101661707)
--(axis cs:10,52.9874298178579);

\addplot [thick, color0]
table {%
1 11.5416805977575
2 12.9866782947381
3 16.5506711260453
4 21.1567937694935
5 187.799247647107
6 318.652177651196
7 362.774451486406
8 389.383714361914
9 409.372567585138
10 418.159412701967
};
\addplot [thick, color1]
table {%
1 9.28993273612333
2 9.29008651499362
3 9.29019350489295
4 9.2902863107349
5 9.29027769984553
6 9.2903521564552
7 9.29039797847755
8 9.29040958121309
9 9.29036252921854
10 9.29039596651179
};
\addplot [thick, color1, dashed]
table {%
1 10.9607062532753
2 16.058512577209
3 24.9977783357691
4 40.1705254893647
5 46.3354882192443
6 51.3588505134873
7 51.1909029125731
8 48.9147681168146
9 48.8395908029938
10 50.4918918750115
};
\addplot [thick, color1, densely dotted]
table {%
1 48.3189650162453
2 43.5197373076435
3 49.3214197893672
4 51.1523921662961
5 46.2133484082181
6 51.8000152441702
7 50.6074666188198
8 49.6802098544906
9 49.2754663928434
10 49.2363199920143
};
\end{axis}

\end{tikzpicture}

%% file: img/burst_stdev/avg_delay.tex
\begin{tikzpicture}

\definecolor{color0}{rgb}{0.12156862745098,0.466666666666667,0.705882352941177}
\definecolor{color1}{rgb}{0.83921568627451,0.152941176470588,0.156862745098039}

\begin{axis}[
width=\fwidth,
height=\fheight,
legend cell align={left},
legend style={fill opacity=0.8,
            draw opacity=1,
            text opacity=1,
            at={(0.45,1.05)},
            anchor=south,
            draw=white!15!black,
            /tikz/every even column/.append style={column sep=0em}},
legend columns=4,
log basis x={10},
tick align=outside,
tick pos=left,
x grid style={white!69.0196078431373!black},
xlabel={$\rho$},
x label style={at={(axis description cs:0.5,-0.25)},anchor=north},
xmajorgrids,
minor x tick num=10,
xminorgrids,
extra x ticks={0.2}, 
extra x tick labels={}, 
extra tick style={tickwidth=\pgfkeysvalueof{/pgfplots/minor tick length}}, 
xmin=0.000767270499010925, xmax=0.260664264112613,
xmode=log,
y grid style={white!69.0196078431373!black},
ylabel={Avg. Delay [ms]},
ymajorgrids,
ymin=0, ymax=70,
ytick style={color=black}
]
\path [draw=color0, thick]
(axis cs:0,20.2501081117748)
--(axis cs:0,24.8059276567894);

\path [draw=color0, thick]
(axis cs:0.001,20.2501081117748)
--(axis cs:0.001,24.8059276567893);

\path [draw=color0, thick]
(axis cs:0.002,20.2501081117748)
--(axis cs:0.002,24.8059276567893);

\path [draw=color0, thick]
(axis cs:0.005,20.1562081845845)
--(axis cs:0.005,24.7609010290602);

\path [draw=color0, thick]
(axis cs:0.01,19.4082334684955)
--(axis cs:0.01,23.8189082345295);

\path [draw=color0, thick]
(axis cs:0.02,19.1188114134613)
--(axis cs:0.02,23.0312255732802);

\path [draw=color0, thick]
(axis cs:0.05,20.5935784804024)
--(axis cs:0.05,22.6353246233468);

\path [draw=color0, thick]
(axis cs:0.1,21.2887289652261)
--(axis cs:0.1,22.1764352036276);

\path [draw=color0, thick]
(axis cs:0.2,22.9531418383693)
--(axis cs:0.2,23.8566588492971);

\path [draw=color1, thick]
(axis cs:0,9.65199393810161)
--(axis cs:0,9.65212388631087);

\path [draw=color1, thick]
(axis cs:0.001,9.65200096026395)
--(axis cs:0.001,9.65212726619007);

\path [draw=color1, thick]
(axis cs:0.002,9.65199393810161)
--(axis cs:0.002,9.65212388631086);

\path [draw=color1, thick]
(axis cs:0.005,9.65199393810161)
--(axis cs:0.005,9.65212388631086);

\path [draw=color1, thick]
(axis cs:0.01,16.588651619629)
--(axis cs:0.01,20.1659716906787);

\path [draw=color1, thick]
(axis cs:0.02,24.9401855516059)
--(axis cs:0.02,31.0885810146148);

\path [draw=color1, thick]
(axis cs:0.05,37.0518844833048)
--(axis cs:0.05,41.8356752621323);

\path [draw=color1, thick]
(axis cs:0.1,41.9028393811769)
--(axis cs:0.1,45.069240121827);

\path [draw=color1, thick]
(axis cs:0.2,47.8943969404069)
--(axis cs:0.2,50.0469928862208);

\path [draw=color1, thick]
(axis cs:0,38.9038647293174)
--(axis cs:0,45.9262065883127);

\path [draw=color1, thick]
(axis cs:0.001,38.9038647293174)
--(axis cs:0.001,45.9262065883127);

\path [draw=color1, thick]
(axis cs:0.002,38.9038647293174)
--(axis cs:0.002,45.9262065883127);

\path [draw=color1, thick]
(axis cs:0.005,38.9038647293174)
--(axis cs:0.005,45.9262065883127);

\path [draw=color1, thick]
(axis cs:0.01,38.731861878773)
--(axis cs:0.01,46.8628050685549);

\path [draw=color1, thick]
(axis cs:0.02,40.2640116800048)
--(axis cs:0.02,48.1227567960265);

\path [draw=color1, thick]
(axis cs:0.05,40.4777185244915)
--(axis cs:0.05,44.7879729757575);

\path [draw=color1, thick]
(axis cs:0.1,42.792629762143)
--(axis cs:0.1,45.5922418425327);

\path [draw=color1, thick]
(axis cs:0.2,47.5021249934674)
--(axis cs:0.2,49.2723719053527);

\path [draw=color1, thick]
(axis cs:0,48.2947946752014)
--(axis cs:0,55.8210372304456);

\path [draw=color1, thick]
(axis cs:0.001,48.2947946752014)
--(axis cs:0.001,55.8210372304456);

\path [draw=color1, thick]
(axis cs:0.002,48.2947946752014)
--(axis cs:0.002,55.8210372304456);

\path [draw=color1, thick]
(axis cs:0.005,48.2947946752014)
--(axis cs:0.005,55.8210372304456);

\path [draw=color1, thick]
(axis cs:0.01,49.8892278779111)
--(axis cs:0.01,58.3826706724407);

\path [draw=color1, thick]
(axis cs:0.02,54.749495042086)
--(axis cs:0.02,62.8763608173807);

\path [draw=color1, thick]
(axis cs:0.05,67.7661209472578)
--(axis cs:0.05,74.0392252610586);

\path [draw=color1, thick]
(axis cs:0.1,86.9869421423416)
--(axis cs:0.1,95.8141491155953);

\path [draw=color1, thick]
(axis cs:0.2,123.191871260789)
--(axis cs:0.2,141.025023854732);

\addplot [thick, color0]
table {%
0.0000001 22.5280178842821
0.001 22.5280178842821
0.002 22.5280178842821
0.005 22.4585546068224
0.01 21.6135708515125
0.02 21.0750184933707
0.05 21.6144515518746
0.1 21.7325820844268
0.2 23.4049003438332
};
\addlegendentry{CBAP Only}
\addplot [thick, color1]
table {%
0.0000001 9.65205891220624
0.001 9.65206411322701
0.002 9.65205891220624
0.005 9.65205891220623
0.01 18.3773116551539
0.02 28.0143832831104
0.05 39.4437798727186
0.1 43.4860397515019
0.2 48.9706949133139
};
\addlegendentry{SP Config. \#1}
\addplot [thick, color1, densely dotted]
table {%
0.0000001 52.0579159528235
0.001 52.0579159528235
0.002 52.0579159528235
0.005 52.0579159528235
0.01 54.1359492751759
0.02 58.8129279297333
0.05 70.9026731041582
0.1 91.4005456289685
0.2 132.108447557761
};
\addlegendentry{SP Config. \#2}
\addplot [thick, color1, dashed]
table {%
0.0000001 42.415035658815
0.001 42.415035658815
0.002 42.415035658815
0.005 42.415035658815
0.01 42.797333473664
0.02 44.1933842380156
0.05 42.6328457501245
0.1 44.1924358023379
0.2 48.38724844941
};
\addlegendentry{SP Config. \#3}
\end{axis}

\end{tikzpicture}